\documentclass{article}

\usepackage{PRIMEarxiv}

\usepackage[utf8]{inputenc} 
\usepackage[T1]{fontenc}    
\usepackage{hyperref}       
\usepackage{url}            
\usepackage{booktabs}       
\usepackage{amsfonts}       
\usepackage{nicefrac}       
\usepackage{microtype}      
\usepackage{lipsum}
\usepackage{fancyhdr}       
\usepackage{graphicx}       
\graphicspath{{media/}}     

\title{SimSched: A Tool For Simulating AUTOSAR Implementations in Simulink
}

\author{
  Jian Chen, Thomas R. Dean \\
  the Department of Electrical and Computer Engineering \\
  Queen's University \\
  Kingston, ON\\
  \texttt{\{jian.chen, tom.dean\}@queensu.ca} \\
   \And
  Manar H. Alalfi \\
  the Department of Computer Science \\
  Toronto Metropolitan University \\
  Toronto, ON\\
  \texttt{manar.alalfi@cs.torontomu.ca} \\
  \And
  Ramesh S \\
  GM R\&D \\
  Warren, MI\\
  \texttt{ramesh.s@gm.com} \\
}

\begin{document}
\maketitle

\begin{abstract}
AUTOSAR (AUTomotive Open System ARchitecture) is an open industry standard for the automotive sector. It defines the three-layered automotive software architecture. One of these layers is the application layer, where functional behaviors are encapsulated in Software Components (SW-Cs). Inside SW-Cs, a set of runnable entities represents the internal behavior and is realized as a set of tasks. To address AUTOSAR's lack of support for modeling behaviors of runnables, languages such as Simulink are employed. Simulink simulations assume Simulink block behaviors are completed in zero execution time, while real execution requires a finite execution time. This timing mismatch can result in failures to detect unexpected runtime behaviors during the simulation phase. This paper extends the Simulink environment to model the timing properties of tasks. We present a Simulink block that can schedule tasks with non-zero simulation times. It enables a more realistic analysis during model development.
\end{abstract}


\section{Introduction}
\section{Introduction}
\label{intro}
Modern automotive systems are software-intensive, and the complexity of these systems is growing. Critical functions of modern vehicles rely on software. An example is embedded controllers that coordinate to perform advanced control functions such as autonomous driving, active safety, and infotainment. All of these functions indicate cars contain some of the most significant pieces of software\footnote{http://spectrum.ieee.org/transportation/systems/this-car-runs-on-code}. A modern high-end car features approximately 100 million lines of code. The worldwide development partnership AUTOSAR \cite{Autosarparternship} was formed to address the challenges of automotive systems. AUTOSAR standardizes the automotive electronic software architecture, and development methodology \cite{Autosarmethodology}.

While many tools support the AUTOSAR process, MATLAB/Simulink (ML/SL) is a popular option for developing automotive software that meets the AUTOSAR standard, providing a toolchain that supports the AUTOSAR development process. The developer can directly use Simulink blocks to create AUTOSAR software components. Embedded Coder\footnote{https://www.mathworks.com/products/embedded-coder.html} provides the mapping of the Simulink models to AUTOSAR components and generates AUTOSAR-compliant production code.

Simulation is a process of representing the actions of a real-world system. Through simulation, engineers can evaluate the system design and diagnose problems early in the design process. However, the Simulink simulation algorithm does not take every factor of the real world into account, such as the time of real-world computation. In other words, software tasks are completed in zero execution time in the simulation stage. In the real world, software tasks take non-zero execution time that varies according to the hardware platform. Hence, simulation cannot reflect an accurate execution at runtime on a specific target hardware platform. Developing a more realistic model of an AUTOSAR-based software application in Simulink is needed.

AUTOSAR supports a modified version of the priority ceiling scheduling\cite{Liu1990}. In this approach, when a lower-priority process uses a shared resource, and a higher-priority process needs access to the shared resource, the priority of the lower-priority process is raised to a higher priority to finish using the resource.  The priority ceiling scheduling is not always the desired behavior. Ideally, any contention over the shared resource should be minimized during the modeling phase. However, this requires an accurate simulation of the timing of each of the software components.

Automotive Electronic Control Unit (ECU) software consists of multiple functions that are often encapsulated in time-triggered tasks and executed on a Real-Time Operating System (RTOS). The use of model-based development in creating the ECU software is limited in that a thread in an ECU is derived from multiple Simulink/Stateflow models that are independently developed, validated, and generated into production code. The concept of thread and timing are mostly absent at the time of development and validation of the models. In addition, the speed of the ECU will be different than the speed on which the simulation is run. Thus there is a large discrepancy between the runtime semantics and the models giving rise to additional work at runtime.  This can be avoided if the runtime abstractions of time-triggered tasks can be captured early at the modeling level. This enables explicit concurrency and timing analysis early in the cycle, thereby reducing the overall time and efforts involved in the development and validation cycle of ECU. In this research, we propose an approach that can reflect the real system behaviors during the simulation phase and in the future, identify race conditions at the model level.

A vital part of SimSched is to make temporal correctness explicit in real-time system development; both logical and temporal correctness is crucial to the correct system functionality. The temporal correctness depends on the timing assumptions of each task.  Our tool SimSched considers the task and runnable timing assumptions, and thereby the temporal logic can be verified during the simulation phase. We have applied this tool to examples to show the ability to identify temporal problems. Our tool results are comparable with other tools such as TrueTime \cite{Henriksson2003}, TRES \cite{Cremona2015}, and our tool is more straightforward to use without writing additional code. Our tool exposes the impact of real-time execution on the semantics of model simulation in the context of AUTOSAR. 

This paper builds on our previous paper presented at the 14th European Conference on Modelling Foundations and Applications \cite{chen2018modeling}. In addition to the description of modeling AUTOSAR implementations in Simulink, we introduce a new graphical user interface (GUI) called SimSched. We also add a model transformation allowing SimSched to subdivide a runnable into subsystems automatically. We compare SimSched with other state-of-the-art tools.

\subsection{AUTOSAR}

AUTOSAR aims to meet the needs of future cars and provides an open industry standard between suppliers and manufacturers. The best way to achieve this goal is to minimize the coupling of software modules through abstraction. AUTOSAR defines three main layers: the application, the runtime environment (RTE), and the basic software (BSW) layer \cite{Autosar}. 

\begin{figure}
	\resizebox{0.99\textwidth}{!}{%
		\includegraphics{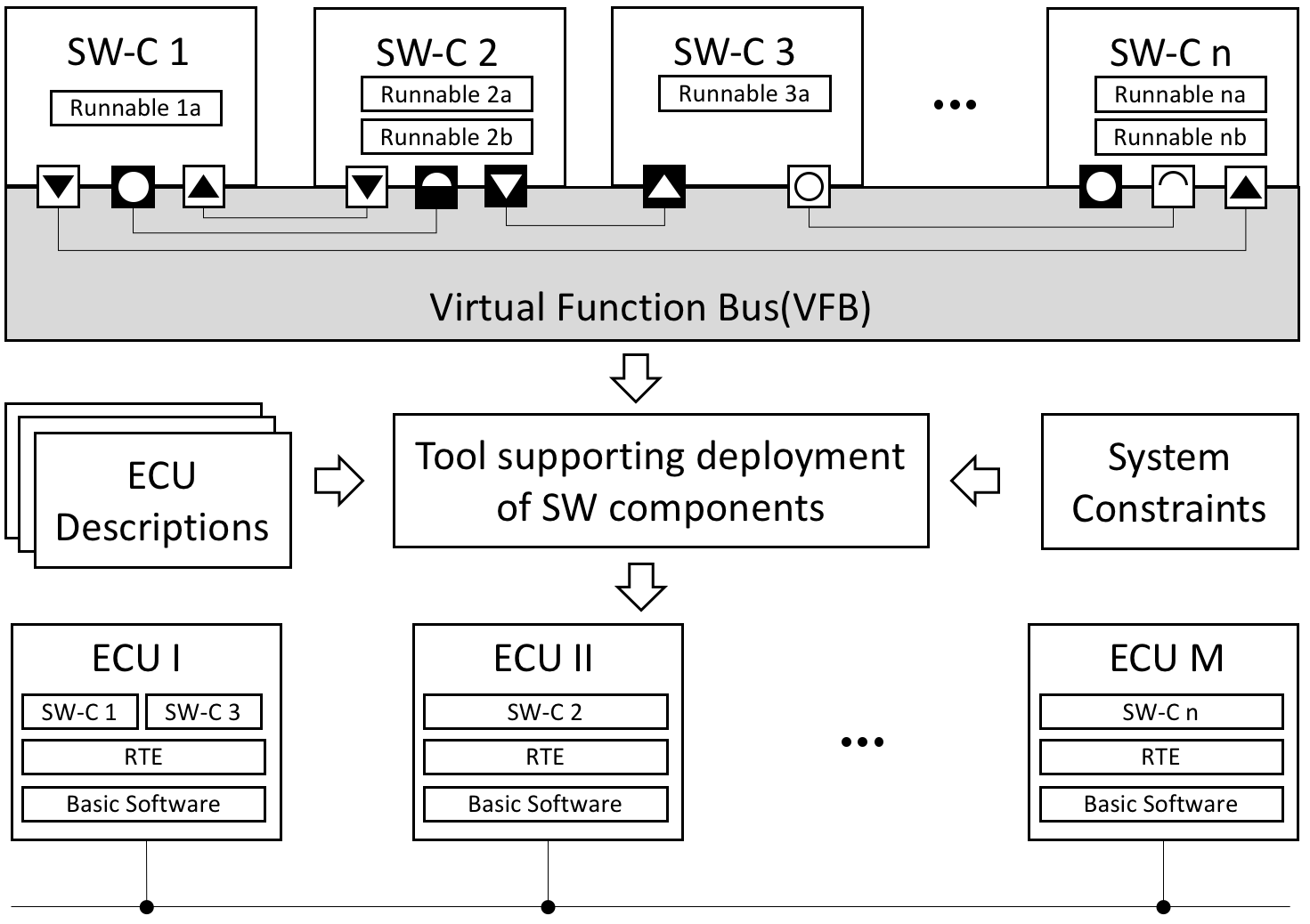}
	}
	\caption{AUTOSAR components, interfaces, and runnables. (Adapted from \cite{Autosarparternship})}
	\label{fig:autosar}       
\end{figure}

The functions in the application layer are implemented by SW-Cs, which encapsulate all or part of the automotive electronic functions, as shown in Figure \ref{fig:autosar}. The components communicate via a Virtual Functional Bus(VFB), which is an abstraction of all the communication mechanisms of AUTOSAR. Using VFBs, engineers abstract the communication details of software components. Inside the SW-Cs,  the internal behaviors are represented by a set of runnables. A runnable is the smallest piece of code that can be independently scheduled, either by a timer or an event. Finally, runnables are combined into a set of tasks on a target platform. Runnables from different components may be mapped into the same task and must be mapped in such a way that ordering relations and causal dependencies are preserved. Generally speaking, the execution order of the runnables inside a task is specified by the engineers according to data-flow dependencies and precedence constraints.

Most of the safety-related automotive applications are realized in real-time applications that combine electronics and software. The temporal execution of applications must satisfy the specific timing requirements to ensure temporal correctness.  Thus, AUTOSAR extended its standards to specific timing concepts in the Specification of Timing Extensions \cite{Autosartimingextension}.  The timing extensions present a formal timing model to define timing requirements and analyze the temporal behavior of a system. In addition, it allows users to define timing constraints and requirements on different abstraction levels. In reality, engineers require development tools to develop systems that satisfy the timing requirements and validate the temporal behavior. In general, to verify the timing properties of real-time applications via simulation- and monitoring-based approaches such as chronSim \cite{chronsim}, RTA-TRACE \cite{rtatrace}. In this research, we attempt to bring the timing specification to the Simulink model to validate the system timing requirements.


\subsection{AUTOSAR Support in ML/SL}
ML/SL has supported AUTOSAR-compliant code generation since version R2006a\footnote{https://www.mathworks.com/products/simulink.html} . ML/SL and Embedded Coder provide a powerful platform for AUTOSAR software development from behavior modeling to production code generation. ML/SL has blocks that represent all AUTOSAR concepts. Existing ML/SL blocks can be used in AUTOSAR development, and no additional AUTOSAR-specific blocks are required. Table \ref{tbl:mapping} shows key mappings between AUTOSAR concepts and Simulink concepts \cite{Applysingsimulink}. ML/SL provides a \emph{Simulink-AUTOSAR Mapping Explorer} for configuring the mapping of Simulink inports, outports, entry-point functions, data transfers, and lookup tables to AUTOSAR elements. Finally, Embedded Coder software supports generating AUTOSAR-compliant C code and exporting AUTOSAR XML(ARXML) description files from an ML/SL model.

\begin{table} 
	\centering
	\caption{AUTOSAR to ML/SL Concept Mapping.}
	\begin{tabular}{ l l}
		\hline
		AUTOSAR & ML/SL   \\ \hline
		Atomic Software Component & Subsystem \\
		Runnable & Function-call subsystem  \\
		RTEEvents & Function calls \\
		\hline
	\end{tabular}
	\label{tbl:mapping}
\end{table}

\subsection{Simulink}
ML/SL system models are blocks connected by signals between input and output ports. The ML/SL simulation engine determines the execution order of blocks before simulation, called the block invocation order. The block invocation order can be determined by the data dependencies among the blocks. ML/SL uses two kinds of block direct feedthrough and non-direct feedthrough to ensure the simulation can follow the correct data dependencies. A block for which its input ports directly determine the output ports is a direct-feedthrough block such as Gain, Product, and Sum blocks. A block for which inputs only affect its state is a non-direct feedthrough block such as a Constant block and Memory block. There is no input port for the Constant block, and the Memory block's output is dependent on its input in the previous time step. ML/SL uses the following two basic rules to form the sorted order \cite{Simulinkguide}: A block must be executed before any of the blocks whose direct-feedthrough ports it drives; Blocks without direct feedthrough inputs can execute in arbitrary order as long as they precede any block whose direct-feedthrough inputs they drive. All blocks are scheduled in sorted order and executed in sequential execution order. The Simulink engine maintains a virtual clock to execute each ordered block at each virtual time. Hence, a Simulink block is usually exhibited as a zero execution time behavior.

Simulink Coder\footnote{https://www.mathworks.com/products/simulink-coder.html} not only supports code generation for ML/SL models, but it also offers a framework to execute the generated code in a real-time environment. The framework assures the generated code follows the simulation engine's standard, and the implementation should preserve the semantics of models. Simulink Coder has two code generation options for periodic tasks: single tasks and multiple tasks. Single-task implementations can preserve the semantics during the simulation because a simple scheduler invokes the generated code in a single thread without preemptions. For multi-task implementations, the generated code is invoked by a rate monotonic (RM) \cite{Lehoczky1989} scheduler in a multithreaded RTOS environment, where each task is assigned a priority and preemptions occur between tasks.

As a consequence of preemption and scheduling, semantic implementation can conflict with the model semantics in a multi-rate system. Hence, the Simulink simulation does not always reflect the actual model behaviors in implementation.  In this work, we develop a model scheduler that can schedule the executions of ML/SL blocks with priorities and preemptions during the simulation.

\subsection{Scheduler}
A set of basic rules determines Simulink blocks' execution order during the updating phase of the simulation. For example, if a block drives other blocks' inputs, it must be executed before it drives. However, the order of execution of Simulink subsystems can be scheduled explicitly in a model.  ML/SL uses a scheduler mechanism to schedule the execution of Simulink subsystems in a specific order explicitly \cite{Stateflow}. Stateflow charts implement the scheduling, and it implicitly controls the order of execution in a Simulink model. Three kinds of schedulers can be implemented using Stateflow, including Ladder logic scheduler, Loop scheduler, and Temporal logic scheduler. These schedulers schedule tasks under the assumption of zero execution time of tasks, which is not reasonable in reality. In this work, we developed a new model scheduler to replace the ML/SL scheduler to enable a more realistic simulation.

\section{Related Work}
Logical Execution Time (LET) \cite{Henzinger2001} was introduced as part of the time-triggered programming language Giotto. It abstracts from the physical execution of a real-time program to eliminate I/O execution time so that a LET model execution is independent of its actual execution. LET uses ports to define a logical task execution, input ports take values at the start of a task, and the output ports release the values at the end of the task execution. LET has an assumption that actual task execution should be able to be finished during the logical execution. Derler \emph{et al.} \cite{Derler2010} demonstrate that real-time software based on LET paradigm could exhibit a similar behavior on a specific platform during the simulation phase in ML/SL. However, Naderlinger \emph{et al.} \cite{Naderlinger:2009:SRS:2349508.2349528} points out that data dependency problems may occur when simulating LET-based software.

Data inconsistencies can be caused by concurrent access to shared data. In order to keep data consistency and to preserve semantics, Ferrari \emph{et al.} \cite{Ferrari2009} discuss the proof of the absence of interference, disabling of preemption, communication buffers, and semaphores as possibilities on a single-core resource in the context of AUTOSAR. Zeng \emph{et al.} \cite{Zeng2011} present similar mechanisms for the preservation of communication semantics for a multi-core platform.

TrueTime \cite{Henriksson2003} simulator is an ML/SL-based network simulation toolbox, and it is suitable for the co-simulation of scheduling algorithms, control algorithms, and network protocols. TrueTime is designed as a research tool that requires a learning curve for system engineers to use this tool. Additionally, tasks cannot be expressed directly using production code and need a particular format for function code.

Cremona \emph{et al.} \cite{Cremona2015} propose a framework TRES, which is used to co-simulate the software model and the hardware execution platform. It adds the schedulers and tasks to Simulink models to model the scheduling delays.

Sundharam \emph{et al.}  \cite{Sundharam2017} apply a  co-simulation approach to tackle the runtime delays problem based on a model interpretation engine Cyber-Physical Action Language (CPAL) \cite{Navet2015}. Tasks are modeled in CPAL, and this CPAL model is added as a Simulink block to study the control performance. Their method does not support any preemptive scheduling policies yet.

Brandberg \emph{et al.} \cite{Brandberg2018} demonstrate a  co-simulation approach to tackle the runtime delays problem based on a model interpretation engine Cyber-Physical Action Language (CPAL) \cite{Navet2015}. Tasks are modeled in CPAL, and this CPAL model is added as a Simulink block to study the control performance. Their method does not support any preemptive scheduling policies yet.

Recently, Naderlinger \cite{Naderlinger2017} introduces timing-aware blocks into ML/SL, which consumes a finite amount of simulation time, so that the simulation behavior of ML/SL models is equivalent to real-time execution behavior.

Our work differs in that we bring the impact of real-time execution to the semantics of model simulation in the context of AUTOSAR. Hence, our approach natively supports AUTOSAR development in ML/SL, and the model scheduler can be integrated into code generation.

\section{Model Scheduler: SimSched}
To reflect the real-time execution of an AUTOSAR Simulink model on actual hardware during the simulation process, we propose a customized scheduler SimSched. It schedules the order of execution of each subsystem at a specific time so that Simulink simulation can capture the real behavior of AUTOSAR applications.

\begin{figure}
		\centering
		\resizebox{0.8\textwidth}{!}{%
			\includegraphics{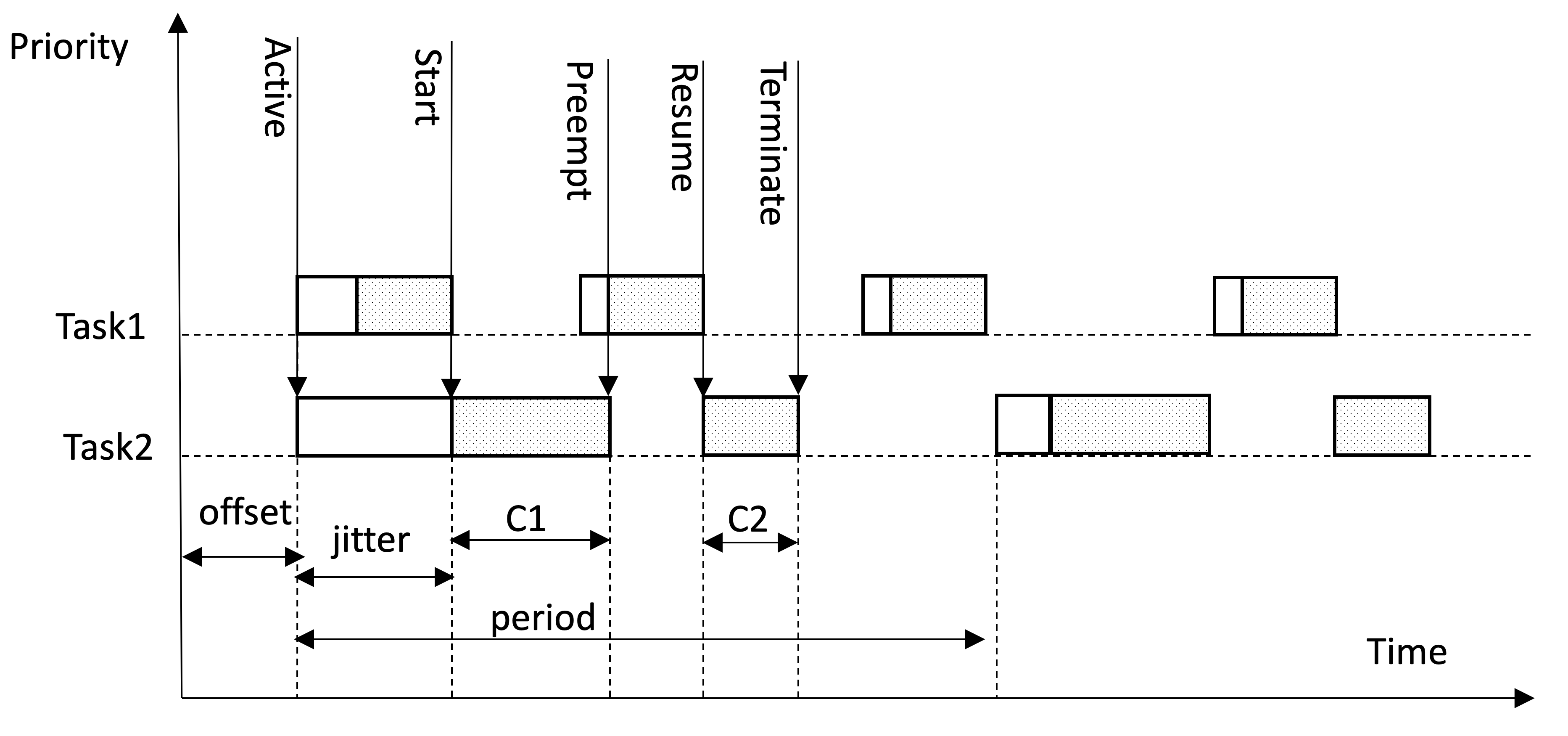} 
		}
		\caption{Task timing parameters shown in Gantt chart (all related to Task2).}
		\label{fig:taskTimingParameter}
	\end{figure}

\emph{Task Model} In automotive software, Simulink models are often created from real-time specifications and are realized as a set of tasks running on an RTOS. In order to better test this kind of software in the Model-in-the-Loop (MIL) phase,  model-based testing needs to be scaled to the real-time context, which includes a timed formalism to model the system under test, and it conforms with the real-time requirements. We define a task model to model the timing properties of tasks in the Simulink environment and the application is modeled as a set of periodic tasks. A task model, $T$, is represented by a tuple $\{\phi, \rho, c, \gamma, prect, precr, prio, jitter\}$, where $\phi$ is an offset of the task, $\rho$ is the period of the task, $c$ is the  Worst Case Execution Time (WCET) of the task,  $\gamma$ is a list of runnables that belong to the task, $prect$ is the precedence constraint of the task, $precr$ is the precedence constraint of the runnables within the task, $prio$ is the priority associated with the task, and $jitter$ is the deviation of a task from the periodic release times. Every task has an implicit deadline which means the deadline of a task is equal to $\rho$. An offset $\phi$ refers to the time delay between the arrival of the first instance of a periodic task and its release time. A WCET is the summation of each runnable execution time. A precedence constraint $prect$ is a list of tasks that specifies the task execution order, $precr$ is a list of runnables within a task. Figure \ref{fig:taskTimingParameter} shows the timing parameters of a task model and  different timing parameters can alter the application's real-time behavior within a system.	

Our model scheduler replaces the Stateflow temporal logic scheduler in the ML/SL model. It schedules a set of given tasks with non-zero execution time so that the model can have real-time behavior during simulation.  The model scheduler is implemented as an S-Function block that can easily substitute for the Stateflow scheduler in an ML/SL model. The model scheduler takes tasks and runnables information as input parameters and outputs scheduled subsystem function call triggers. Inside the model scheduler, we implemented a preemptive scheduling algorithm written in C based on the Fixed Priority Scheduling (FPS) algorithm \cite{Liu1973}, which computes the scheduling, and the model scheduler outputs a subsystem function call trigger when a task is scheduled. A function call trigger is a control signal, which triggers the connected subsystem to execute when a control signal has a function-call event.

While one of the standard scheduling algorithms in OSEK/AUTOSAR is priority ceiling scheduling, often the developer would prefer to minimize the task priority changes due to shared resources. In our approach, we use FPS so that we can identify race conditions that occur in the model that would require a priority inversion. In FPS, each task has a fixed priority preassigned by the developer, and the tasks are stored in a ready queue in an order determined by their priorities. The highest priority task is selected from the ready queue to execute. The oldest task will be chosen if there are more than one of the same priority tasks exist. Suppose a higher-priority task is scheduled during the execution of a lower-priority one in a preemptive system. In that case, the more top-priority task is executed immediately, and the lower-priority task is moved to the ready queue. The RM scheduling algorithm is one of the widely used FPS, and it is used in Simulink Coder for code generation. In RM scheduling, the priority of a task is associated with its period; if a task has a smaller period,  then it has a higher priority.

The S-Function provides a mechanism to extend the capabilities of ML/SL by customizing blocks, and S-Functions can be accessed from a block diagram so-called \emph{S-Function block}. Customized algorithms can be added to Simulink models via S-Function blocks written either in MATLAB or in C. An S-Function block has a parameter field that users pass specify parameters to the corresponding S-Function block. S-Functions communicate with the ML/SL engine through a set of callback functions, the so-called S-Function API. Thus, S-Functions make it possible to control the ML/SL simulation process by using a customized algorithm.

\section{Implementation: SimSched}
SimSched is implemented as a custom configurable Simulink library block that includes an FPS algorithm written in C and invokes the connected subsystem in the ML/SL interactive development environment.

\subsection{Manual Prototype}
\begin{figure}
	\centering
	\resizebox{0.7\textwidth}{!}{%
		\includegraphics{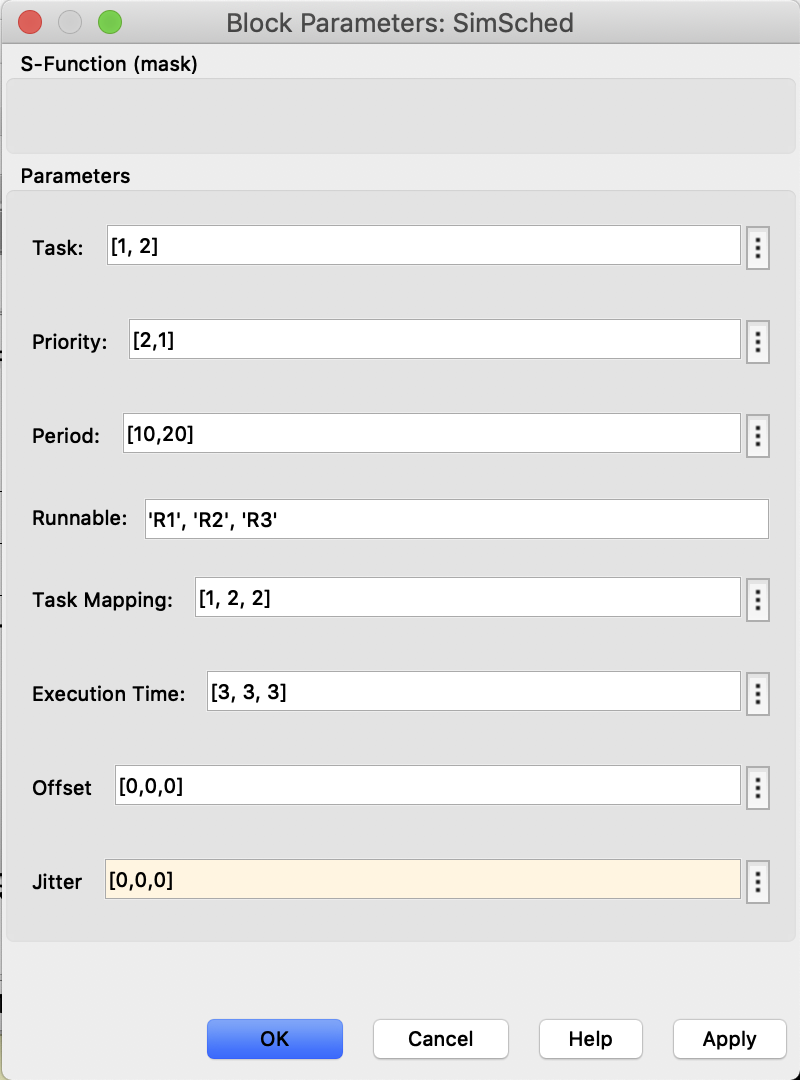}
	}
	\caption{SimSched parameters. The user can enter the corresponding setting parameters according to the specific runnable task mapping situation.}
	\label{fig:msParameter}       
\end{figure}

An S-Function contains a set of callback functions, which interact with the simulation engine and are executed at different stages during a simulation. We use output functions \emph{(mdlOutputs)} that compute the output values based on the input parameters. Before running a simulation, we provide the necessary parameters in the block parameters dialogue shown as in Figure \ref{fig:msParameter}. The parameters include \emph{Task}, \emph{Priority}, \emph{Period}, \emph{Runnable}, \emph{Task Mapping}, \emph{Execution Time}. The user-entered parameters are implemented by a block mask, which provides the parameter dialogue box for users to input.

\begin{figure}
	\resizebox{0.99\textwidth}{!}{%
		\includegraphics{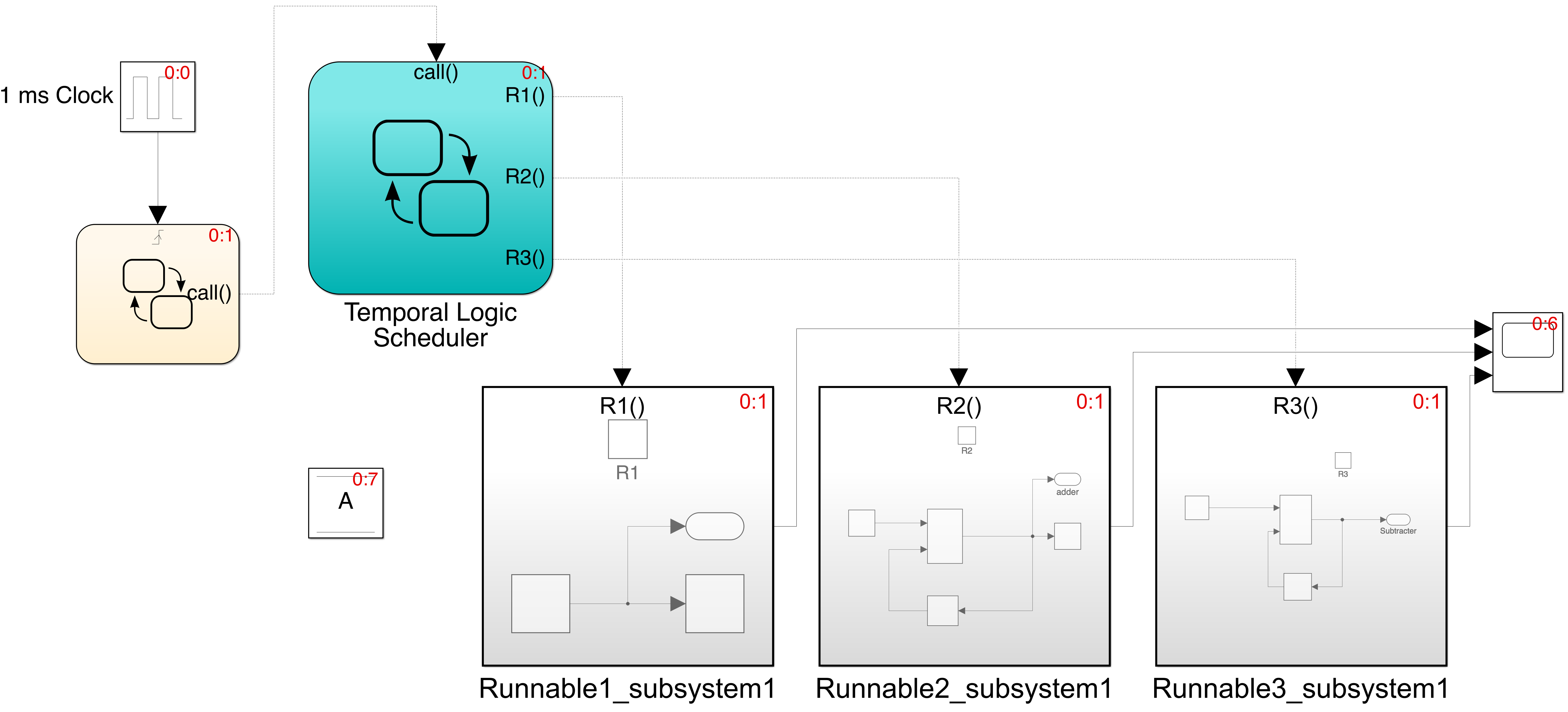} 
	}
	\caption{A simple example of using Stateflow to schedule AUTOSAR SW-Cs with the execution order.}
	\label{fig:stateflowExecOrder}
\end{figure}

Before running a simulation, the ML/SL engine sorts all blocks in an order based on their topological dependencies determined by feedthrough, which specifies a partial order execution. Total order execution is then selected to execute among the partial order execution compatible with the partial order imposed by the model semantics. In general, the execution order of a specific Simulink subsystem is determined by a Stateflow scheduler. Figure \ref{fig:stateflowExecOrder} shows an example of using Stateflow to schedule runnables that explicitly displays the execution order on the top-right corner of each block. The order notation uses two numbers to represent the block order. The first number is the index of a model or a subsystem the block resides in; the second number is the execution order within an execution context. For example, the Runnable1\_subsystem1 has a sorted order 0:1, 0 means it is at the root level of a model, and 1 specifies the sorted order's block position.

\begin{figure}
	\resizebox{0.99\textwidth}{!}{%
		\includegraphics{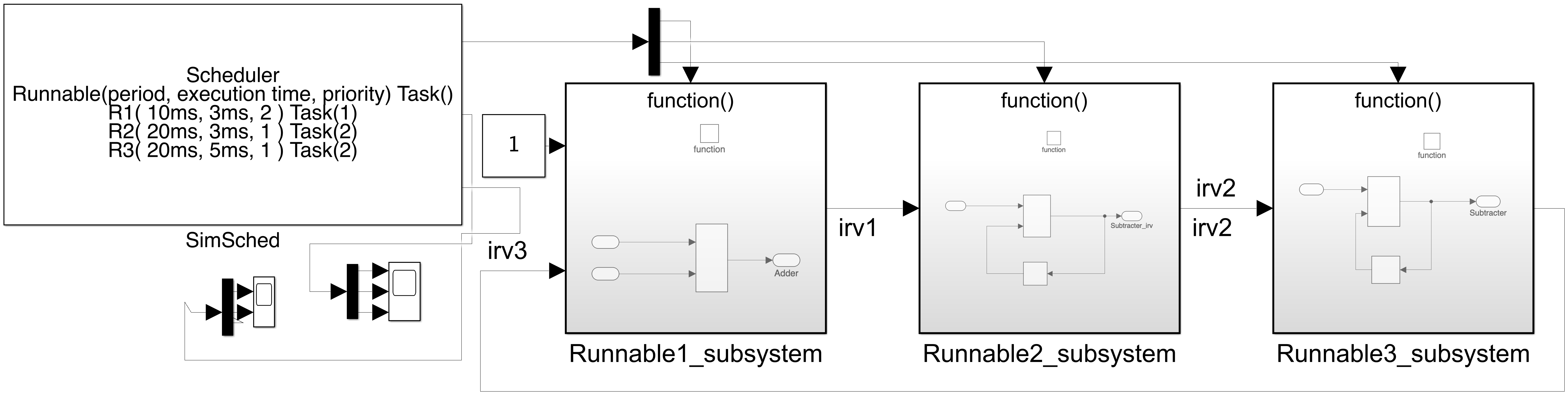} 
	}
	\caption{A simple example of using SimSched to schedule AUTOSAR SW-Cs.}
	\label{fig:msSimpleExample}
\end{figure}

Figure \ref{fig:msSimpleExample} illustrates the use of our model scheduler. In this example, SimSched is at the top left corner, which schedules three runnables, and they are mapped to two tasks. For example, R1 is mapped to Task1, the period for Task 1 is $10ms$, priority is $2$, and the execution time of $R1$ is $3ms$. R2 and R3 are mapped to Task 2. The period for Task 2 is $20ms$, priority is $1$, and the execution time of $R2$ and $R3$ are $3ms$ and $5ms$ accordingly. From Table \ref{tbl:mapping}, we know that each function-call subsystem represents an AUTOSAR runnable. The function-call subsystem can be executed conditionally when a function-call event signal arrives. Both an S-function block and a Stateflow block can provide such a function-call event. Our model scheduler applies the function-call invocation mechanism to use an S-function to generate a function-call event to schedule each runnable (function-call subsystem).

In our case, the Stateflow scheduler is replaced by our tool model scheduler, and each Simulink subsystem is scheduled to be executed at a specified time with a finite execution time. In the previous version of this paper, we manually remove the Stateflow Temporal Logic scheduler,  added our model scheduler, and connected each runnable block to  SimSched orderly according to the original model's execution order. The parameters are manually set through the parameter dialogue box.    Figure \ref{fig:msSimpleExample} shows the result of our manual prototype. SimSched can yield three outputs. The first output port is the trigger signal that periodically outputs the specified time for each subsystem. The trigger signals are connected to a demux block, which splits the multiple trigger signals into a single signal to trigger each subsystem. The second output port is a runnable activation chart that illustrates each runnable schedule. It shows the start and finishes time of each runnable, including the runnable execution time. The third output port is a task activation chart that illustrates each task schedule.

\subsection{Automated Transformation}

\begin{figure}
	\centering
	\resizebox{0.8\textwidth}{!}{%
		\includegraphics{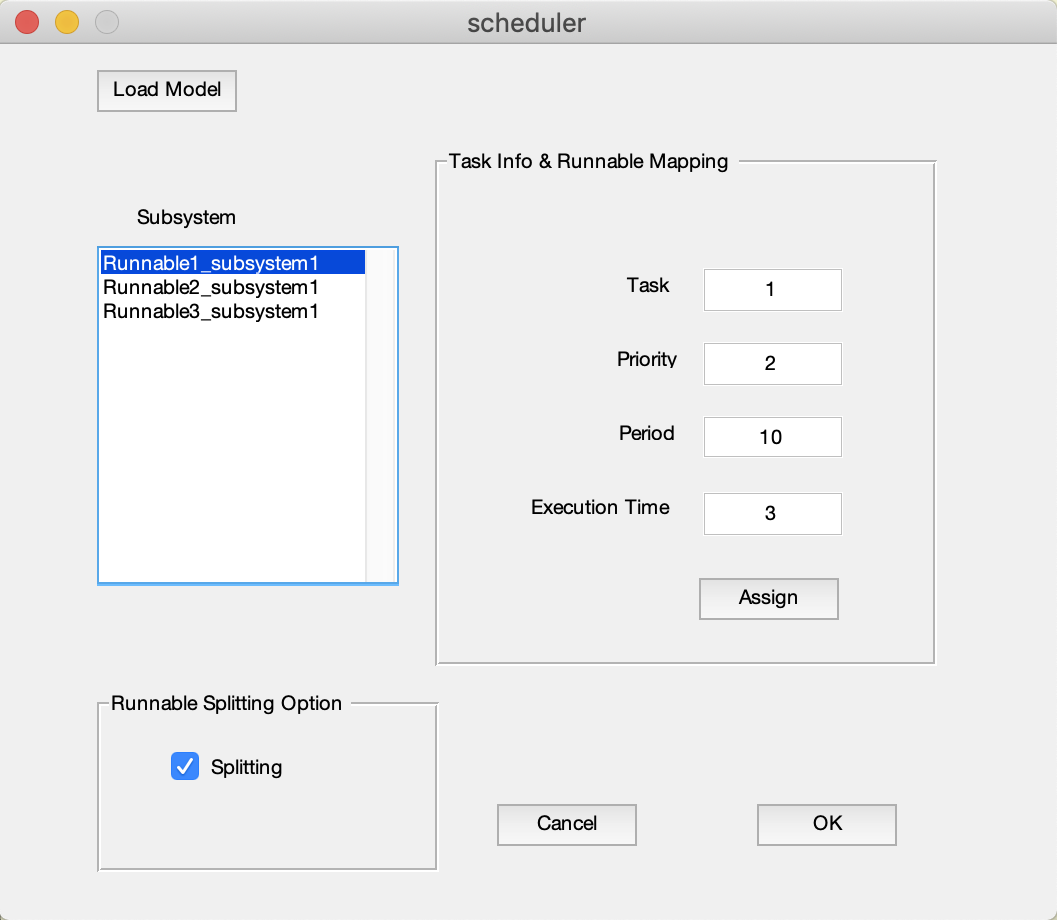} 
	}
	\caption{Graphical user interface of SimSched .}
	\label{fig:schedulerGUI}
\end{figure}

We implemented our tool with a new GUI design and integrated SimSched with the Simulink environment. SimSched is implemented in Matlab without the use of any other third-party libraries so it can be easily integrated into any existing workflow. The GUI, shown in Figure \ref{fig:schedulerGUI}, is designed to ensure that users can easily accomplish the scheduling task.

SimSched has additional functionality, which can automatically identify each runnable block, add the model scheduler block, and add connections from the identified SimSched to the runnables. SimSched also allows users to assign task information to each runnable accordingly via the GUI.

The transformation is implemented in the following steps. First,  SimSched looks for all function-call subsystems using the ML/SL application program interface(API). Usually, runnables are at the root level of a Simulink model, so SimSched uses the API function to find all function-call subsystems at the root level when a developer loads a model file. After finding all function-call subsystems, SimSched displays the names of each identified runnable in a list box. Figure \ref{fig:schedulerGUI} shows that SimSched has identified three runnables in the example and lists them in the GUI list box for the next step. 

\begin{figure}
	\centering
	\resizebox{0.99\textwidth}{!}{%
		\includegraphics{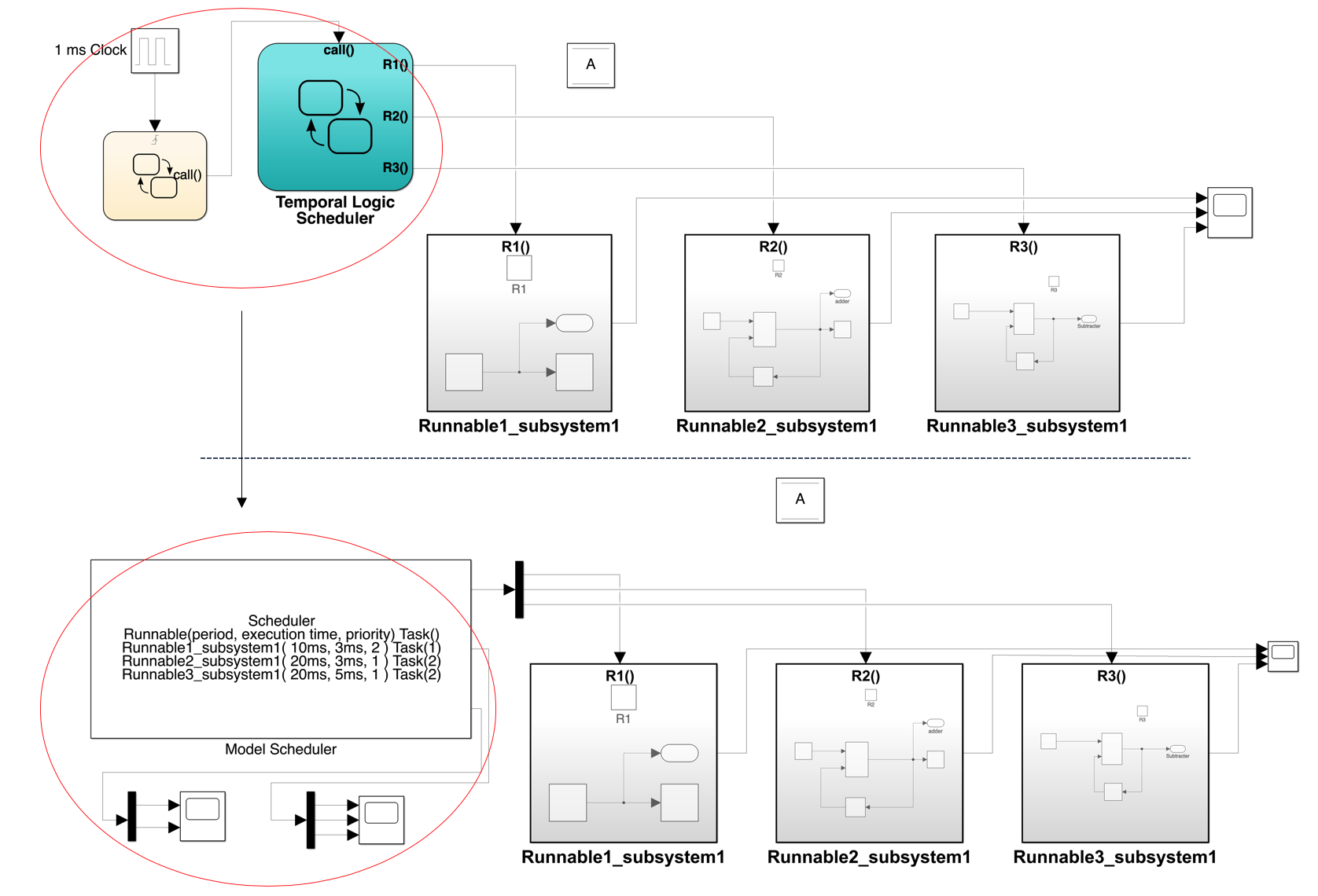} 
	}
	\caption{SimSched is able to connect each runnable automatically.}
	\label{fig:autoScheduler}
\end{figure}

In the second step, the user provides task information, including which task a runnable is assigned to,  the task's priority, and period and execution time information. SimSched maintains a task-runnable mapping table that stores the information for each task and which runnables are mapped to each task. This mapping information is the key for our model scheduler to schedule each runnable. 

Figure \ref{fig:schedulerGUI} shows an example of these parameters.  Runnable1\_subsystem1 is highlighted, showing the parameters for this runnable on the right.  Runnable1\_subsystem1 is assigned to task 1,  the priority of task 1 is 2, the period of task 1 is $10ms$, and the execution time of Runnable1\_subsystem1 is $3ms$.

In the final step, SimSched adds the scheduler block to the Simulink model and connects each of the runnable to the SimSched scheduler block automatically. This step obtains the original model's execution order, deletes the connections between the runnables and the Stateflow scheduler, adds the SimSched block, and connects each runnable to the SimSched block in the same execution order. Figure \ref{fig:autoScheduler} shows the result of adding SimSched and connecting the runnables. We can see the Stateflow scheduler has been disconnected from the figure, and the SimSched block is added and connected to the three function-call subsystems. The parameters of SimSched have been set automatically based on the task-runnable mapping table. 

\subsection{Fine Grain Transformation}
Task preemption can be either fully preemptive or partially cooperative\cite{sanudo2016schedulability,biondi2017logical}. Interruptions of preemptive tasks may occur at any execution point, while interruptions of cooperative only happen at runnable boundaries. The preemption of runnable is not allowed to occur during the execution of a RunnableEntity in cooperative tasks\cite{AutosarRTEsoftware}. In other words, preemptive runnables can be preempted by higher priority runnables; higher-priority cooperative runnables can preempt cooperative runnables at boundaries. To model fully preemptive scheduling, SimSched has a transformation option shown at the left bottom of Figure \ref{fig:schedulerGUI}. With this option, we can simulate the preemption at the runnable level. When this option is selected, the SimSched transformation splits each runnable into multiple function-call subsystems so that the model scheduler schedules each subsystem individually. We name this transformation process a fine-grain transformation. The input of transformation is the original Simulink model, in which function-call subsystems represent runnables. After transformation, each original runnable is transformed into multiple function-call subsystems, and each new function-call subsystem contains one block, which is part of the original runnable.

\begin{figure}
	\centering
	\resizebox{0.6\textwidth}{!}{%
		\includegraphics{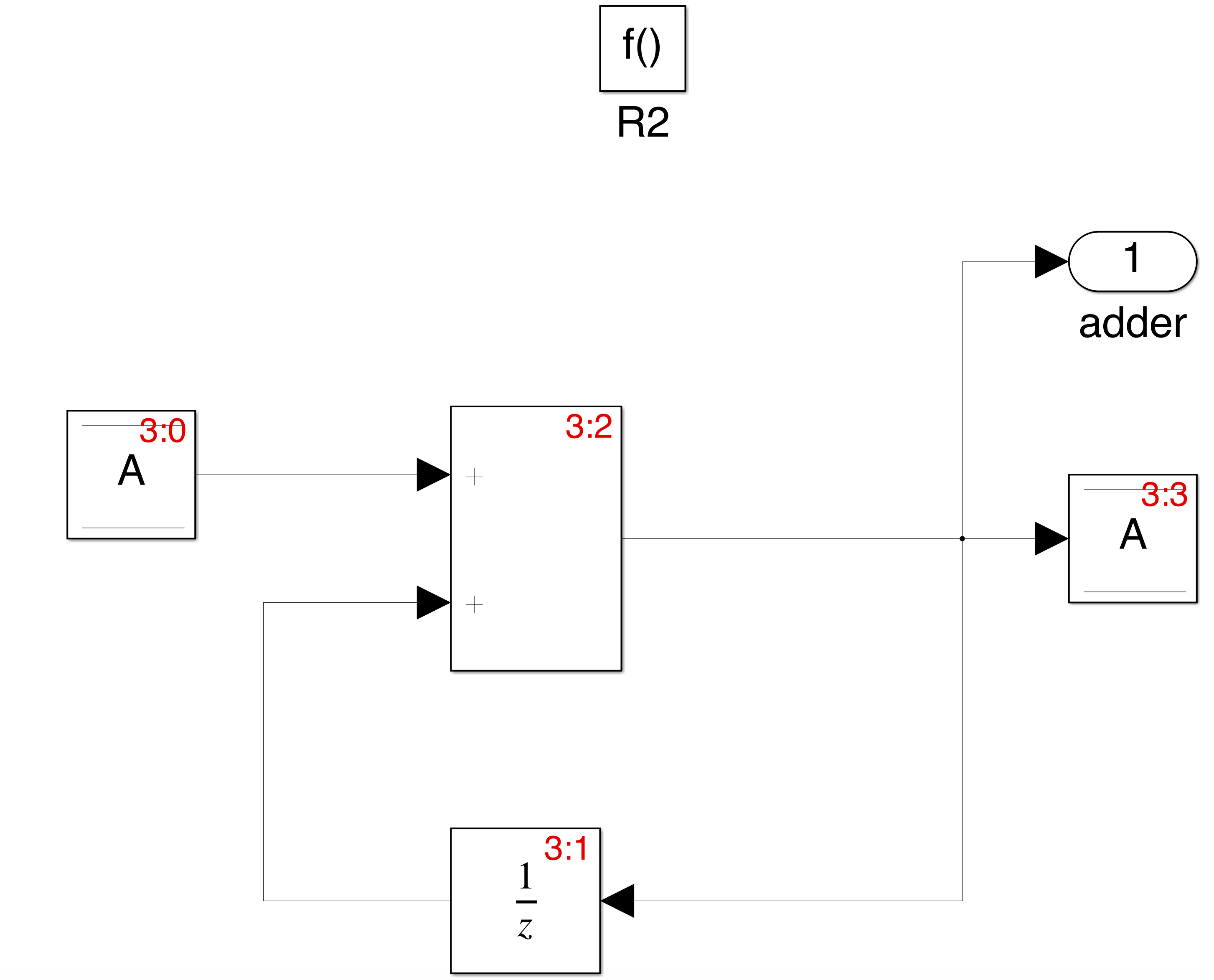} 
	}
	\caption{Runnable2\_subsystem1 with the execution order.}
	\label{fig:sub2execorder}
\end{figure}
SimSched flattens each Simulink subsystem block and puts each of the internal blocks into a new function-call subsystem so that SimSched can schedule it.

First, SimSched identifies all runnables and all blocks within an identified runnable when a user loads a model into SimSched, and also obtains the execution order of each block from the Simulink simulation engine for later use. We use the previous simple example (Figure \ref{fig:stateflowExecOrder}) to explain the model transformation process. When a user selects the model splitting option, SimSched locates all runnables inside the loaded Simulink model and identifies all of the blocks inside each runnable.  In this example, three runnables are recognized first, and each block's execution orders within a runnable are stored. A closer look at Runnable2\_subsystem1 is shown in Figure \ref{fig:sub2execorder}. This subsystem contains six blocks:  function-call trigger port, Data Store Read block, Data Store Write block, Sum block, Unit Delay block, and Outport. The Data Store Read block with the sorted order 3:0 is executed first. The rest blocks execute in numeric order from 3:0 to 3:3. SimSched preserves the same execution order for the transformed model.

In the second step of the transformation, SimSched copies each block to a new function-call subsystem. When adding a block to a function-call subsystem,  SimSched uses a table to store the source ports, destination ports of each block, the handle of the new function-call subsystem block, and each block's execution order. SimSched will name the new function-call subsystem using the pattern: the original subsystem name $+$ is a sequential number. The sequential number is the execution order of the original subsystem. The Data Store Read block is in the first place of the sorted order, so the new function-call subsystem is named R2\_Runnable2\_subsystem1\_1. The second executed block is the Unity Delay block and named R2\_Runnable2\_subsystem1\_2. Both blocks are the input to  R2\_Runnable2\_subsystem1\_3, which contains the Sum block. The last completed block is the Data Store Write block, and it is in the R2\_Runnable2\_subsystem1\_4 subsystem.

\begin{figure}
	\centering
	\resizebox{0.3\textwidth}{!}{%
		\includegraphics{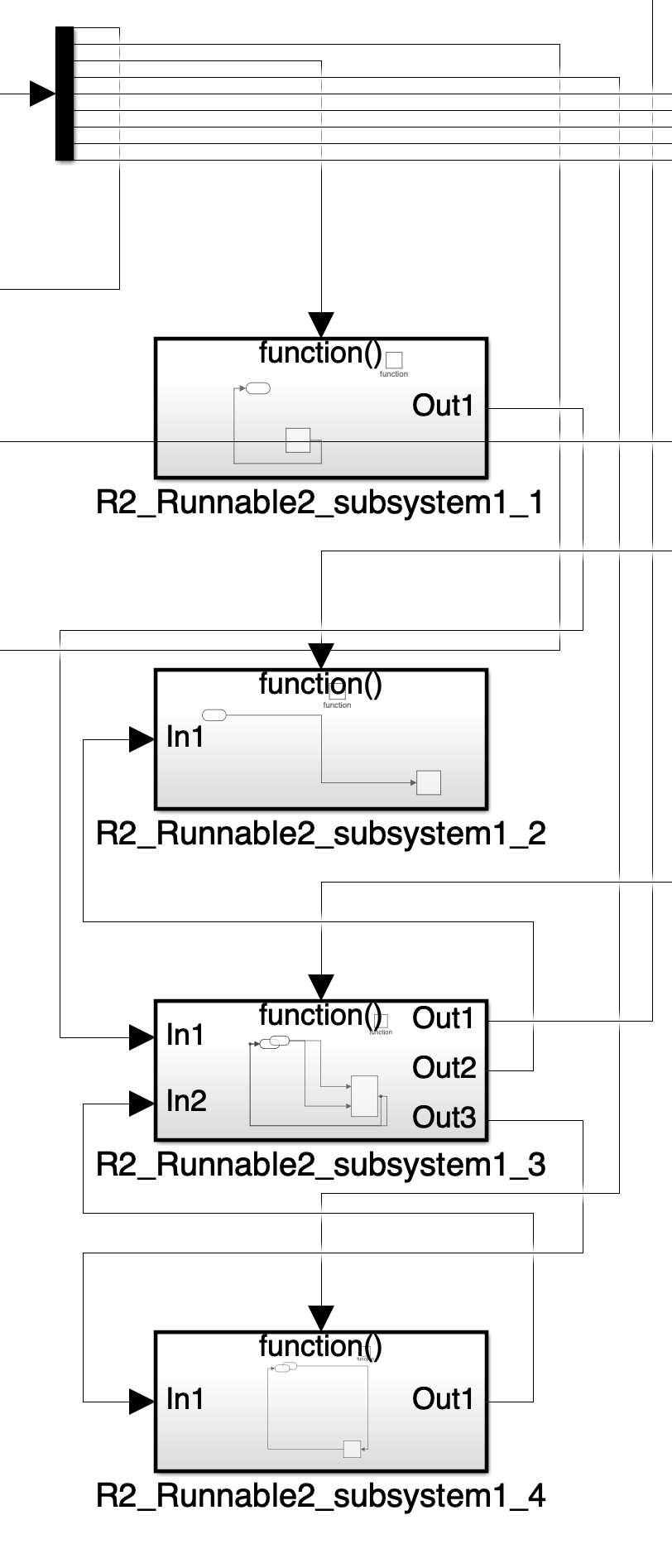} 
	}
	\caption{Runnable2\_subsystem1 after transformation.}
	\label{fig:sub2after}
\end{figure}

In the third step of the transformation, SimSched connects the blocks based on the port connectivity obtained from the previous step. Table \ref{tbl:connectivitytbl} shows an example of the ports connectivity table from Figure \ref{fig:sub2execorder}. It contains four columns and shows the source ports and destination ports of each block. The first column shows the name of a block. The second column shows a handle number of the new function-call subsystem. The third column shows the inport connection. The fourth column shows the outport connection. For example, the first row of Table \ref{tbl:connectivitytbl} shows the Data Read block's name in the first column. The Data Read does not have the inport connection, so the third column is empty. Data Read has an outport, and its destination is Sum block inport1. Each block in the new function-call subsystem still keeps the same connectivity as inside the subsystem. However, the $inports$ and the $outports$ of a subsystem are omitted since they connect directly to the source or destination. SimSched also adds additional an inport or outport to the new function-call subsystem accordingly so that the new function-call subsystem can be connected in the same way as the original model connection. Figure \ref{fig:sub2after} shows the transformation result of Runnable2\_subsystem1. For example, The Sum has two inports and three outports. Now, the R2\_Runnable2\_subsystem1\_3 function-call subsystem contains the Sum block, and we can see it has two inports and three outports. Two inports are connected to the two subsystems: one contains the Data Read block, and the other contains the Unit Delay block. 

\begin{table} 
	\centering
	\caption{Example of intermediary processing connectivity table.}
	\begin{tabular}{ c c c c }
		\hline
		Block & Function-call Block & inport & outport \\
		\hline
		Data Read  & \#handle &  & Sum inport1  \\
		Unit Delay   & \#handle &Sum outport1 & Sum inport2  \\
		Sum   & \#handle & 1:Data Read outport1 &  1:outport1 \\
		Sum   & \#handle & 2:Unit Delay outport1 & 2:Data Write inport1 \\
		Sum   & \#handle & & 3: Unit Delay inport1 \\
		Data Write  & \#handle & Sum outport1 &  \\
		\hline
	\end{tabular}
	\label{tbl:connectivitytbl}
\end{table}

The fourth step is to set model scheduler parameters. SimSched treats each new function-call subsystem as a new ``runnable'' with original task mapping information. Elevated blocks are mapped to the same task as before, and they share the same runnable information. For example, Figure \ref{fig:sub2after}  shows runnable Rununnable2\_subsystem1 has been transformed into four subsystems. Originally, Rununnable2\_subsystem1 is assigned to Task2 with execution time $3ms$. SimSched still assigns the four subsystems to Task 2 with the same priority and period. The four subsystems share an execution time of $3ms$, and each subsystem has an execution time of $0.75ms$.  

The last step is to connect the new blocks to the SimSched block. The connection to the new block strictly follows the original execution order to avoid introducing new abnormal execution. 

SimSched is based on an FPS algorithm and implemented as a level 2 S-function written in C. Our algorithm takes as arguments the six input parameters mentioned above. The parameters can be grouped into two levels. One describes the properties of a task such as \emph{priority} and \emph{Period}; the other one represents the properties of runnables such as \emph{Task Mapping} and \emph{Execution Time}. SimSched reserves this two-level information and computes the current active runnable signal. In this work, we assume the execution time of each runnable is already known. The execution time could either be measured by running the code on a test platform or by analyzing the behavior of generated code (or Simulink model) by off-the-shelf tools. Usually, the FPS algorithm computes a scheduling table, and then a runtime dispatching algorithm invokes each task according to the pre-computed table. 

\subsection{SimSched Scheduling}
Simulating a model has three phases: model compilation, link phase, and simulation loop phase \cite{Sfunction}.  The Simulink engine each time goes through the loop is defined as one simulation step. In each simulation step, SimSched is executed and computes the running task and runnable of current sampling time and yields a signal to the output port when the current sample time is a beginning of a task period. The output signal triggers the connected subsystem, which is runnable of the current execution task. The model scheduler determines the current active runnable and the associated task information at each simulation step. If a task or a runnable is expected to run at this simulation step, then the model scheduler invokes a macro to trigger the subsystem connected to SimSched.

\begin{figure}
	\centering
	\resizebox{0.9\textwidth}{!}{%
		\includegraphics{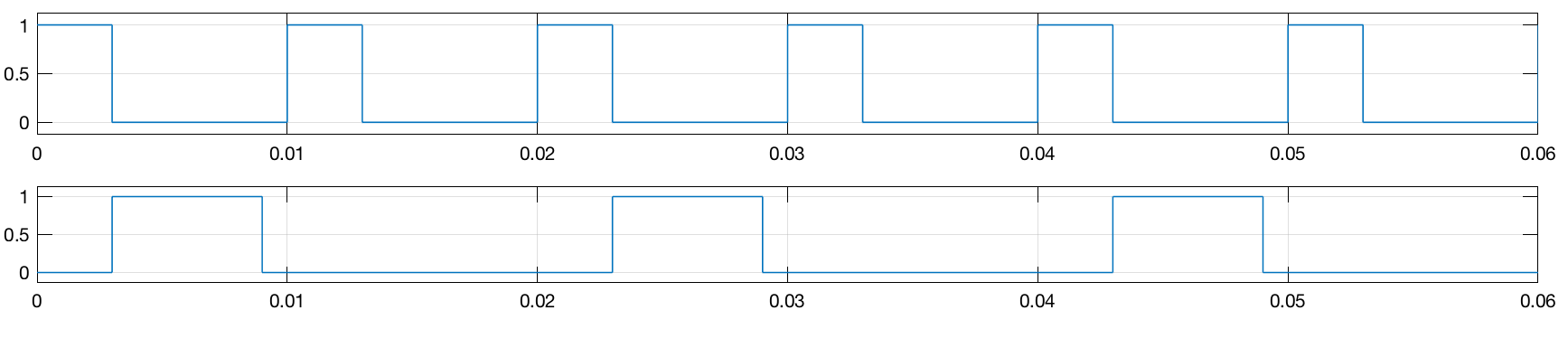} 
	}
	\caption{Task active chart.}
	\label{fig:msSimpleExampleTask}
\end{figure}
\begin{table} [htbp]
	\centering
	\caption{The simple example settings.}
	\begin{tabular}{ c c c c c}
		\hline
		Task & Period($ms$) & Execution & Priority & Runnable \\
		&                       & Time($ms$) &          &                  \\ \hline
		$T_1$   & 10 & 3 & 2 & $R_1$ \\
		$T_2$   & 20 & 3 & 1 & $R_2$ \\
		$T_2$   & 20 & 3 & 1 & $R_3$ \\
		\hline
	\end{tabular}
	\label{tbl:simpleparameter}
\end{table}

We now present an example of how SimSched performs the schedule computation in Figure \ref{fig:msSimpleExample}. The simple example has the following settings shown in Table \ref{tbl:simpleparameter}. During the simulation loop phase, task $T_1$ and $T_2$ are all scheduled at the first simulation step, and SimSched maintains a scheduling table to store the scheduled tasks. $T_1$ is the only executed task at the first simulation step due to its higher priority, and the execution of $T_1$ takes 3ms as only runnable $R_1$ is mapped to $T_1$. In the first simulation step, SimSched output a function-call signal to trigger $T_1$ that is connected to the first output port of demux \emph{Runnable1\_subsystem}. There are no output signals at the simulation step two because it is still during the execution of $T_1$. Until simulation step three, $T_1$ completes its execution and it is time to trigger $T_2$. $R_2$ and $R_3$ are mapped to $T_2$, so they have the same priority and period. $R_2$ is executed at this simulation step because its connection order is before $R_3$. $R_3$ is executed right after the completion of $R_2$. After the execution of $R_3$, that is an idle time, so there is no trigger signal being output. Figure \ref{fig:msSimpleExampleTask} illustrates the task execution process during each simulation step. $T_1$ is active during the first three simulation steps, and $T_2$ is active at the following six simulation steps in the first $10ms$ period.

\section{Demonstration}
This section uses AUTOSAR-compliant  ML/SL models scheduled by our model scheduler to show the scheduler can capture the actual behavior on a hardware platform during a Simulink simulation.

\subsection{AUTOSAR Model Scheduling}
\begin{figure}
	\centering
	\resizebox{0.99\textwidth}{!}{%
		\includegraphics{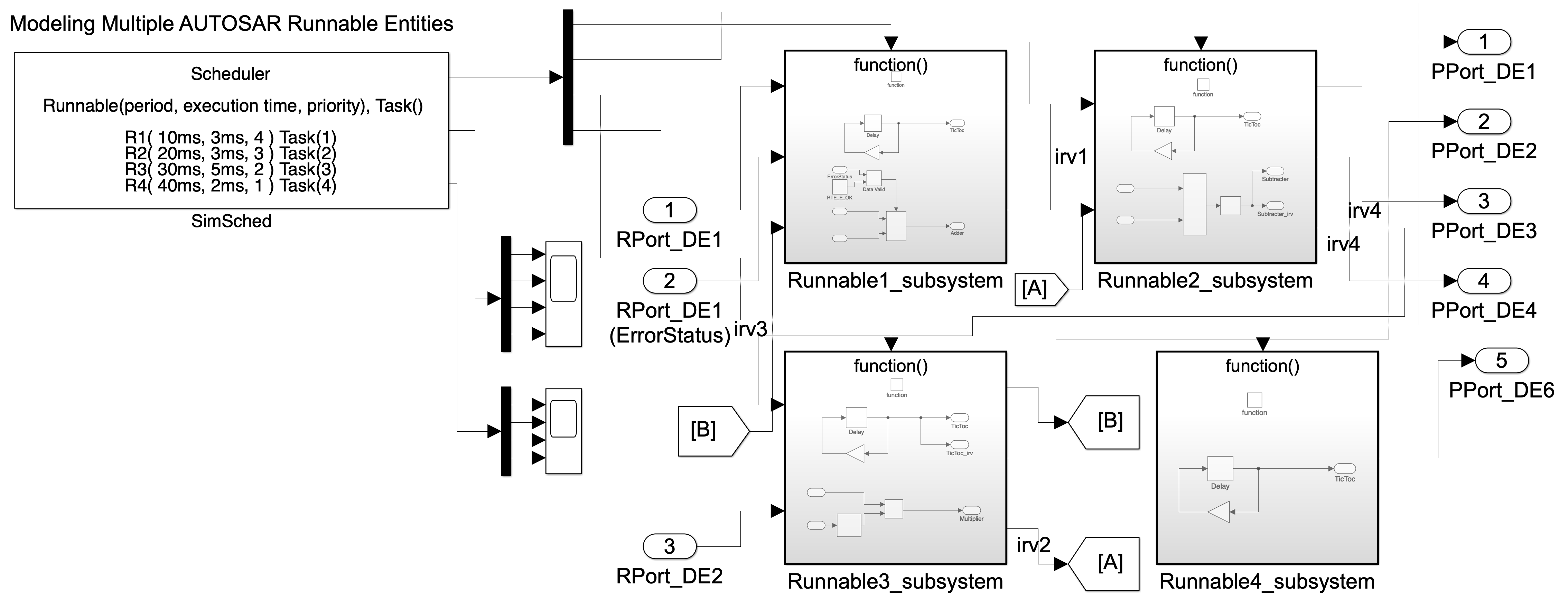} 
	}
	\caption{Using SimSched to schedule AUTOSAR SW-Cs.}
	\label{fig:msExample}
\end{figure}

First, we demonstrate a simple example that shows how Model in the Loop (MIL)  analysis benefits from our model scheduler. Figure \ref{fig:msExample} shows a simple example using a model scheduler to schedule four runnables mapped to three tasks. The details of the settings are shown in Table \ref{tbl:parameter}.

\begin{table} 
	\centering
	\caption{Using model scheduler parameters setting.}
	\begin{tabular}{ c c c c c}
		\hline
		Task & Period($ms$) & Execution & Priority & Runnable \\
		&                       & Time($ms$) &          &                  \\ \hline
		$T_1$   & 10 & 3 & 3 & $R_1$ \\
		$T_2$   & 20 & 3 & 2 & $R_2$ \\
		$T_2$   & 20 & 5 & 2 & $R_3$ \\
		$T_3$   & 30 & 2 & 1 & $R_4$ \\
		\hline
	\end{tabular}
	\label{tbl:parameter}
\end{table}

In this example, we have four runnables which are mapped to three tasks:  $R_1$ is mapped to $T_1$; $R_2$ and $R_3$ are mapped to $T_2$; $R_4$ is mapped to $T_3$. The model scheduler takes the parameters of tasks as input to calculate three outputs. The first output is the runnable triggers, which are connected to four subsystems accordingly. The other two outputs are time execution diagrams of tasks and runnables.

If we simulate this example using a standard scheduler, the execution order of this example is $T_1$$T_2$$T_3$ or $R_1$$R_2$$R_3$$R_4$ respectively and they are completed within the first $10ms$ period based on tasks settings. In reality, the above order is not possible since each task requires a certain amount of execution time on a hardware platform. Our scheduler tries to be realistic by taking execution time into account.

\begin{figure}
	\centering
	\resizebox{0.76\textwidth}{!}{%
		\includegraphics{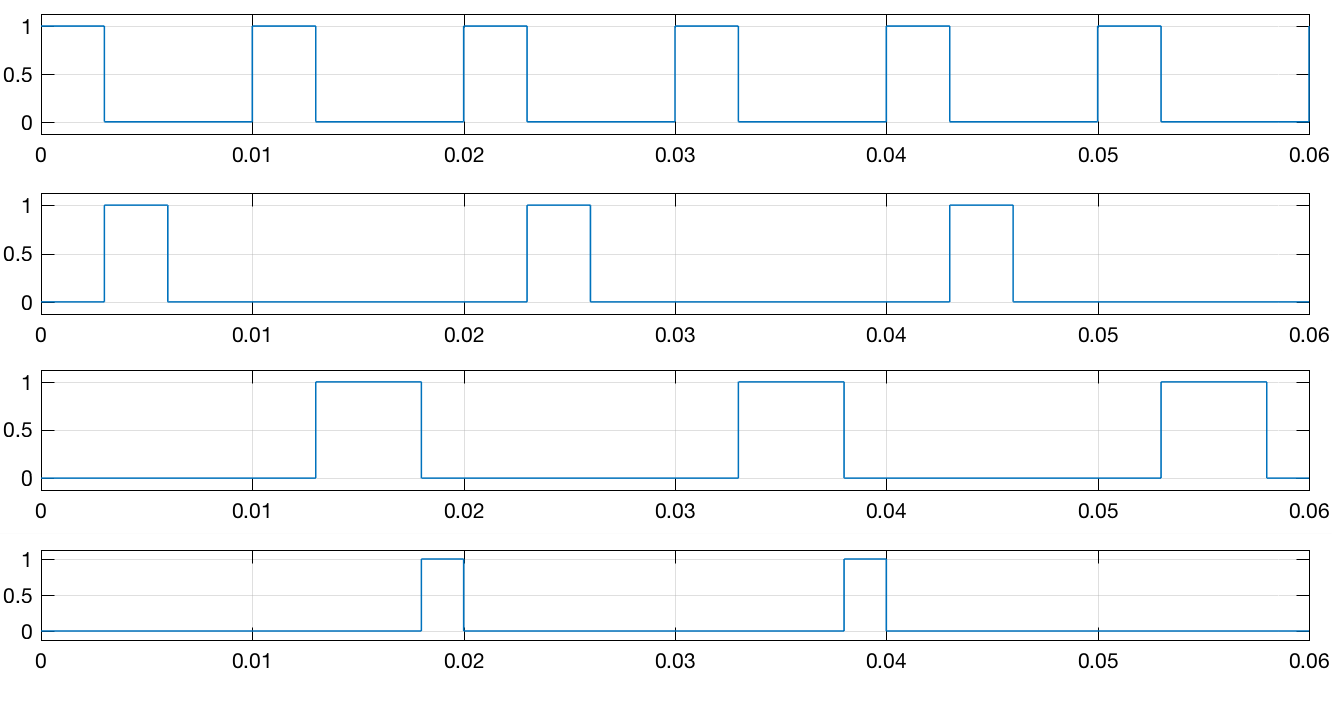} 
	}
	\caption{Runnables execution time diagram.}
	\label{fig:runnableted}
\end{figure}

\begin{figure}
	\centering
	\resizebox{0.76\textwidth}{!}{%
		\includegraphics{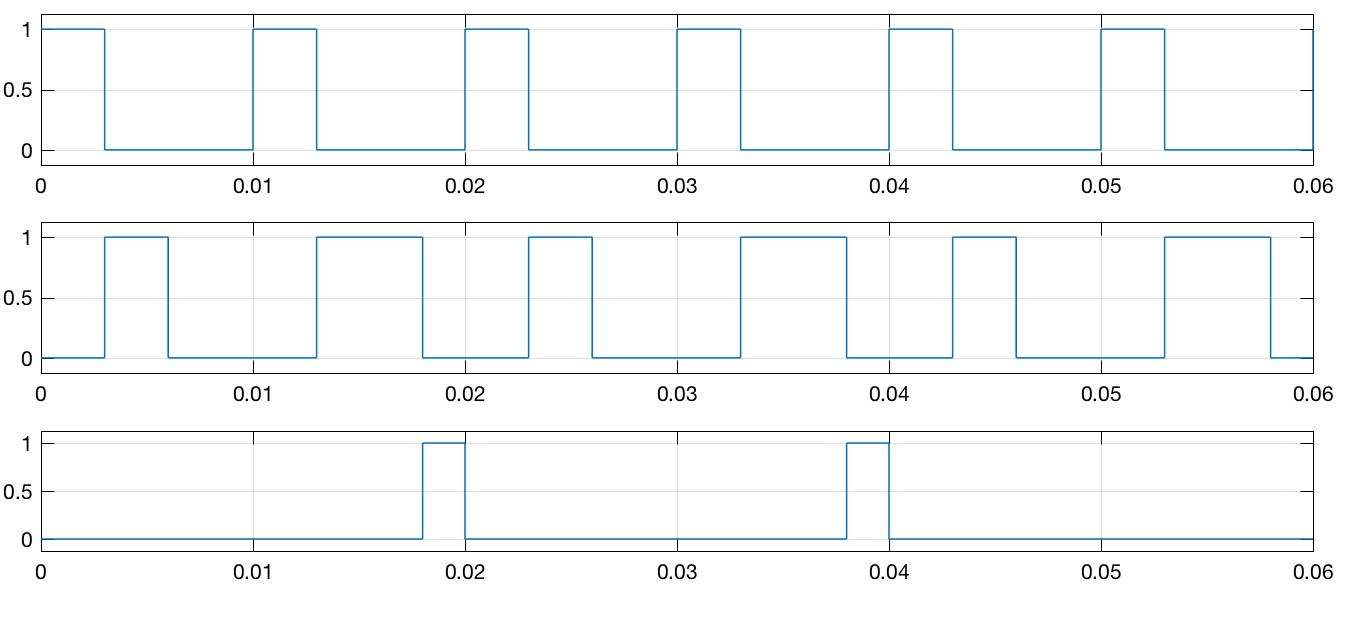} 
	}
	\caption{Tasks execution time diagram.}
	\label{fig:taskted}
\end{figure}

Figure \ref{fig:runnableted} shows the runnables execution diagram of our scheduler. There are four output signals and each output signal represents each runnable. The first signal shows the execution of $R_1$ in $T_1$. It has the highest priority so it is trigged at the beginning and takes $3ms$ execution time. After the execution of $T_1$, the next highest priority is $T_2$ with 2 runnables. $R_2$ is triggered at time of $3ms$ and takes another $3ms$ execution time. $R_3$ is supposed to be triggered right after the completion of $R_2$. However, $R_3$ is triggered right after the completion of the second $R_1$ instance because the period of $T_1$ is $10ms$ and the execution time for both $R_1$ and $R_2$ are $3ms$. There is only $4ms$ left before $R_1$ is triggered at the next period and it is less than the execution time of $R_3$ of $5ms$. Because a runnable is the smallest atomic component within an SW-C,  there is no preemption between runnables. $R_3$ cannot be preempted by $R_1$. Thus, $R_3$ is scheduled to be executed after the completion of the second $R_1$ instance. Our execution model has an assumption that the runnables are atomic units. Hence, the simulated processor is idle between $0.006$ and $0.01$. One could relax this by allowing runnables to be preemptable. The execution order of our scheduler is $T_1$$T_2$$T_1$$T_2$$T_3$ or $R_1$$R_2$$R_1$$R_3$$R_4$. Figure \ref{fig:taskted} shows the tasks execution diagram. There are three output signals that represent task execution. During the first period of $T_2$, $T_1$ is triggered twice and $T_2$ is preempted by $T_1$.

\subsection{Scheduler Example}
This section uses an example model, which is scheduled by two different schedulers, the Stateflow Scheduler and SimSched, to show our scheduler can simulate actual behaviors during the simulation phase. The Stateflow scheduler takes zero execution time during the Simulink simulation. On the contrary, the model scheduler triggers each subsystem, which takes a specified execution time during the simulation. By comparing simulation results between these two schedulers, the potential unexpected behaviors of Simulink models are exposed.

\begin{figure}
	\resizebox{0.99\textwidth}{!}{%
		\includegraphics{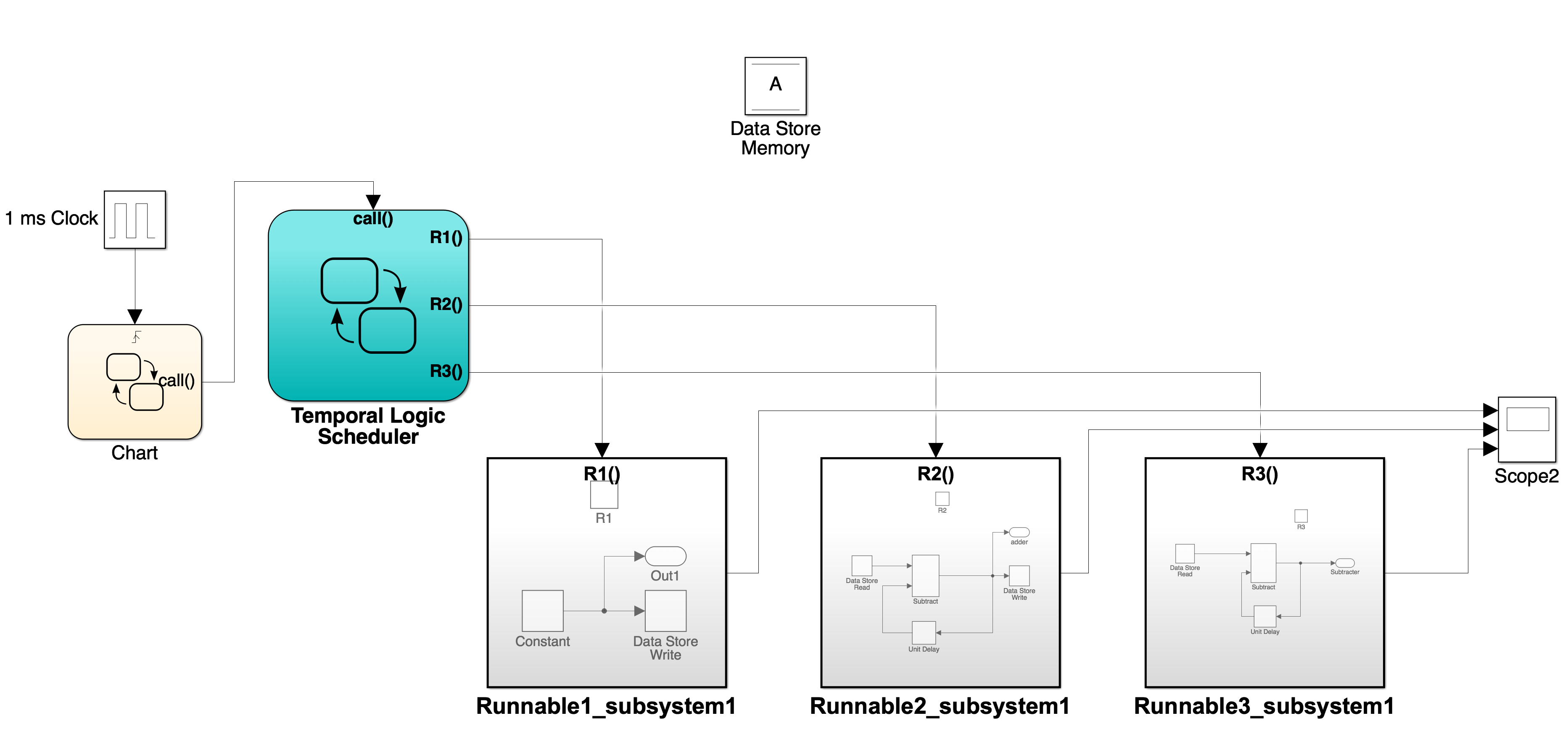} 
	}
	\caption{An example using Stateflow scheduler.}
	\label{fig:normalexample}
\end{figure}

\begin{figure}
\centering
\begin{minipage}{.5\textwidth}
  \centering
  \includegraphics[width=.98\linewidth]{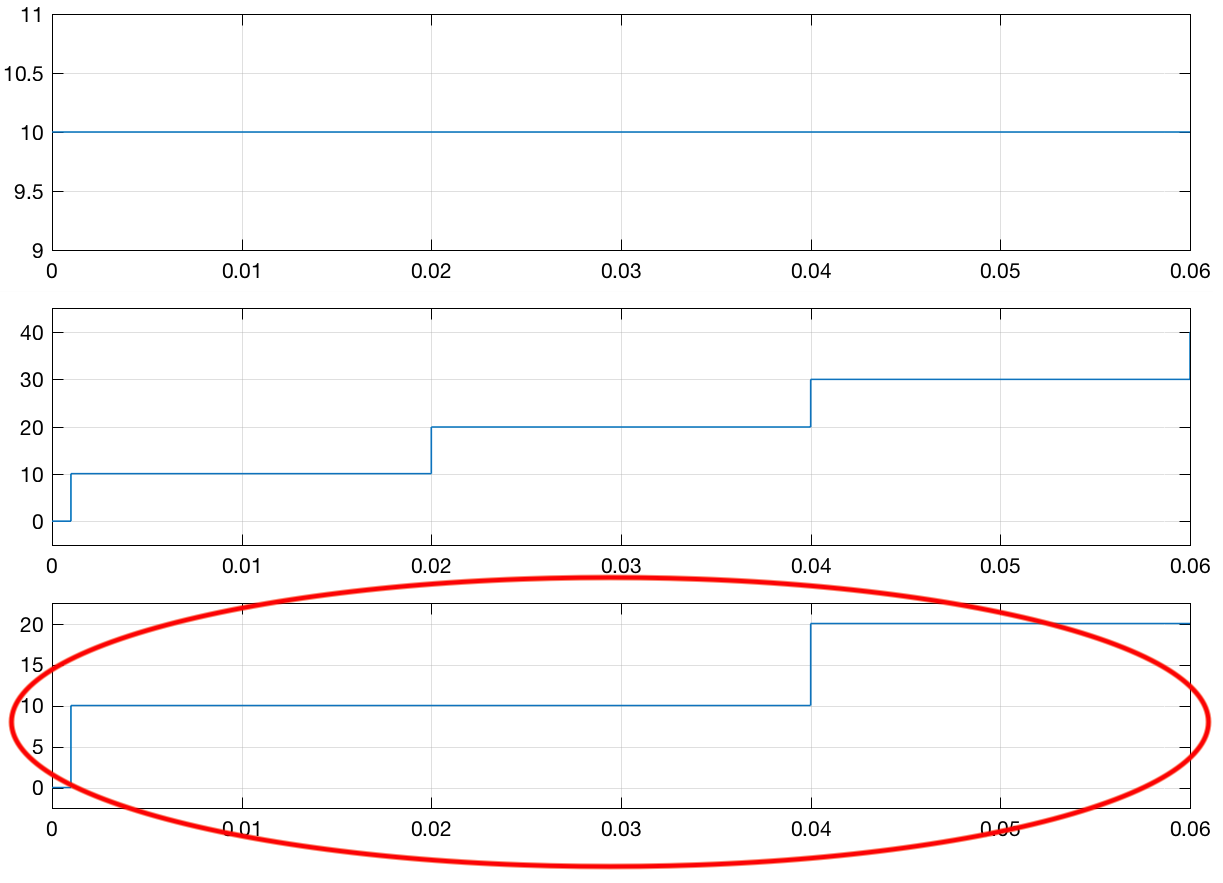}
  \caption{Output of Stateflow scheduler.}
  \label{fig:dataracenormal}
\end{minipage}%
\begin{minipage}{.5\textwidth}
  \centering
  \includegraphics[width=.99\linewidth]{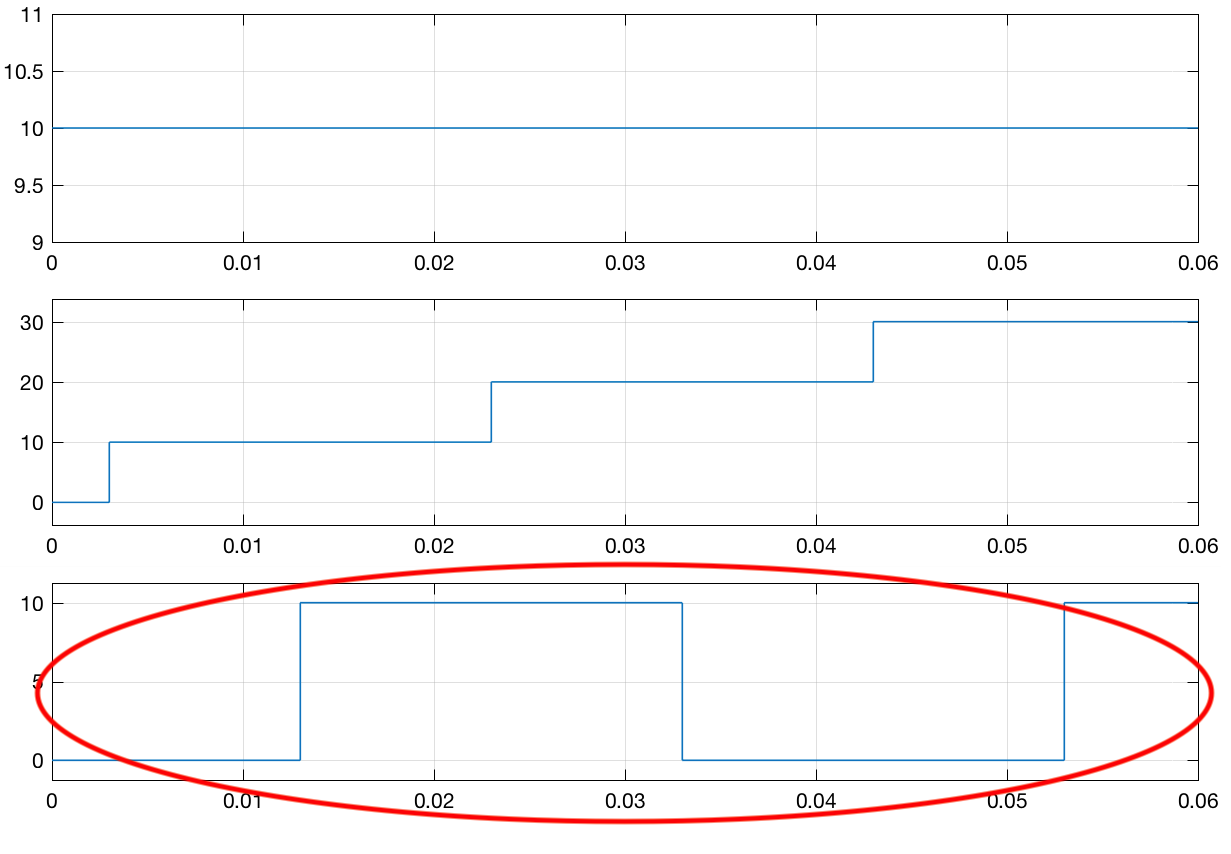}
  \caption{Output of SimSched.}
  \label{fig:dataraceschdeuler}
\end{minipage}
\end{figure}

Figure \ref{fig:normalexample}  shows the normal example which uses a Stateflow temporal logic scheduler to trigger each task. The parameter settings are the same as Table \ref{tbl:simpleparameter} except the execution time of $R_3$ is $5ms$. There are three runnables ($R_1$, $R_2$, $R_3$) mapping to two tasks ($T_1$, $T_2$) in this example. $R_1$ writes a constant value to a global variable $A$. $R_2$ reads $A$ first then writes the summation of $A$ and its delay value to $A$. $R_3$ reads $A$ then subtracts its delay value from $A$, and outputs the result. Figure \ref{fig:dataracenormal} shows the simulation output of this normal example. The three signals are the outputs of $R_1$, $R_2$, and $R_3$ from top to bottom. From this simulation result, the output of $R_3$ is an increasing number. In the normal simulation, the execution orders are $T_1$$T_2$ or $R_1$$R_2$$R_3$.



We replaced the Stateflow scheduler with our model scheduler and run the simulation again, we can get a different result shown in Figure \ref{fig:dataraceschdeuler}. In the second simulation,  the output of $R_3$ is a pattern of zero, constant value, which is different from the previous example. The execution order of this example is $T_1T_2T_1T_2$ or $R_1R_2R_1R_3$. In the previous example, $R_3$ always reads $A$ which is written by $R_2$. Using our scheduler, $R_3$ reads the global variable $A$ from the output of the second $R_1$ instance because $T_2$ is preempted by $T_1$ during the execution of $T_2$.

So far, we have demonstrated the model scheduler is capable of simulating task interference of cooperative tasks. In this new version, we extended our SimSched to be able to schedule preemptive as well. As explained previously, we have implemented a transformation that allows SimSched to schedule at the level of individual blocks. 

\begin{figure}
	\resizebox{0.99\textwidth}{!}{%
		\includegraphics{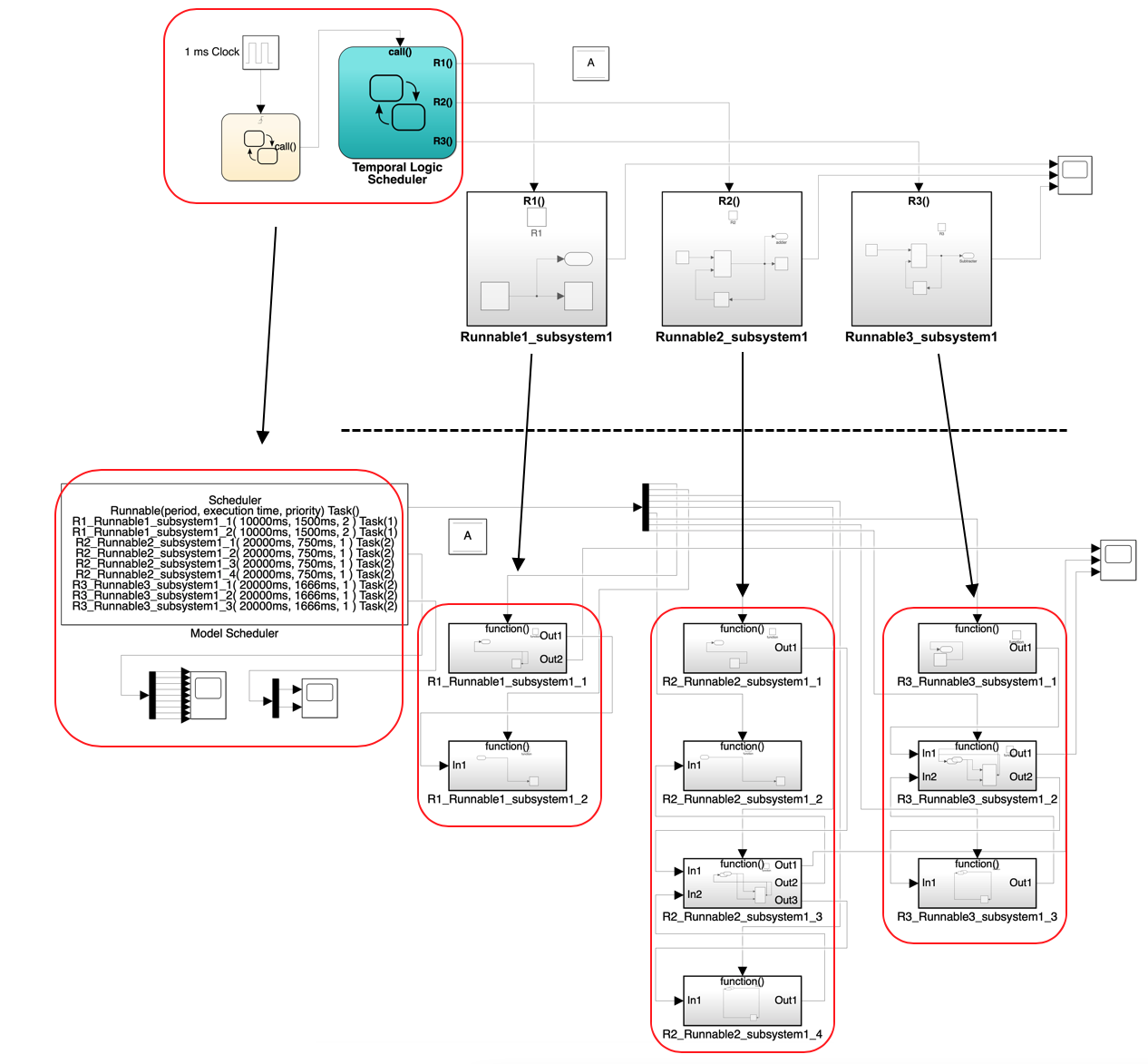} 
	}
	\caption{Splitting runnables into multiple functions calls subsystems automatically.}
	\label{fig:autoSplitting}
\end{figure}

\begin{figure}
	\centering
	\resizebox{0.76\textwidth}{!}{%
		\includegraphics{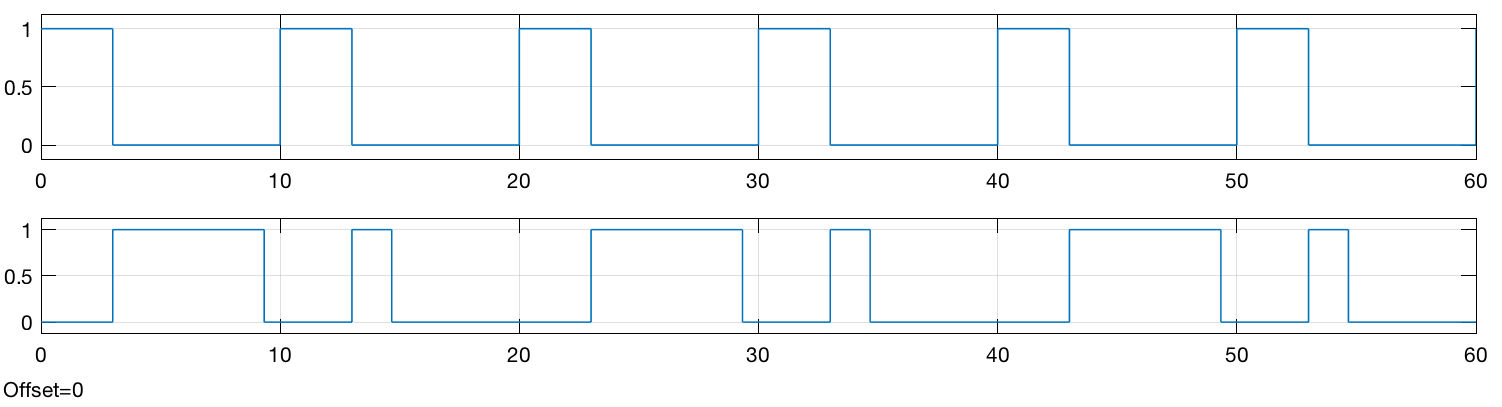} 
	}
	\caption{Tasks execution time diagram after splitting runnables into multiple function call subsystem.}
	\label{fig:splittingTask}
\end{figure}

This fully split example is shown in Figure \ref{fig:autoSplitting}. After the fine-grained transformation, every Simulink block inside the original runnable (function-call subsystem) has been moved to a new function-call subsystem, and it can be treated as a separate runnable. The connections of new subsystems are the same as the original connections. Each new runnable's execution time is shorter because each new runnable contains only a part of the original runnable subsystem. Figure \ref{fig:splittingTask} shows the tasks execution time diagram after splitting the runnables into multiple function call subsystems. In the above example, task $T_1$ now has two runnable, and task $T_2$ has seven runnables. The total execution time of the original runnables is unchanged, and the execution time of each new runnable is computed from the system's total execution time. The use of multiple runnables allows the task preemption point to change. In Figure \ref{fig:splittingTask}, we can see the preemption still occurs between  $T_1$ and $T_2$, but only the subtraction block of original $R_3$ is preempted. $R_3$ still reads $A$ which is written by $R_2$. The output has the same value as the Stateflow scheduler execution. Our SimSched can simulate the preemption between $T_1$ and $T_2$, but this preemption may not cause any problems in the real world. With the ability to simulate preemptions at the runnable level, our SimSched can simulate real-world execution more precisely. 

\subsection{Three servos example }
We can apply SimSched to regular Simulink models other than AUTOSAR models. AUTOSAR models use regular Simulink blocks to represent AUTOSAR concepts, and they do not have any AUTOSAR-specific blocks. Our SimSched can be easily adapted to other Simulink models after altering the scheduling algorithm. 

\begin{figure}
	\centering
	\centering
	\resizebox{0.99\textwidth}{!}{%
		\includegraphics{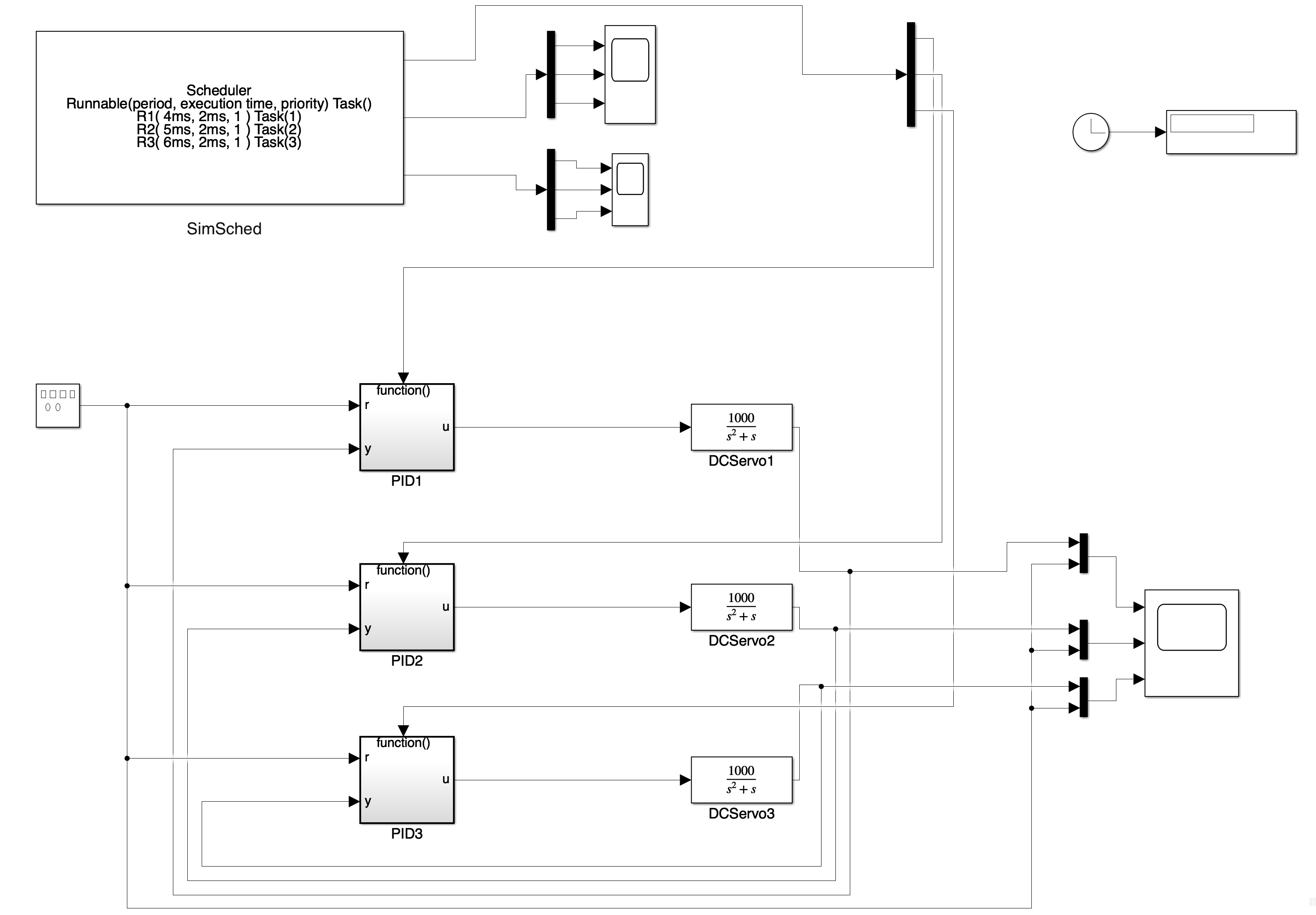} 
	}
	\caption{The three servos example adapted from \cite{Cremona2015}.}
	\label{fig:MSthreeServos}
\end{figure}

We adapt an example from the TrueTime example library, which shows a possible implementation of a three-servo PID control system. This example is also adapted by the TRES and CPAL controller in Simulink. The example is shown in Figure \ref{fig:MSthreeServos}. In this example, three DC servos are modeled by a continuous-time system, and three PID controllers are implemented as three subsystems. We use the same settings as other tools and map three controller subsystems to three runnables $R_1$, $R_2$, and $R_3$ then they are mapped to tasks $T_1$, $T_2$, and $T_3$. The task periods are $T_1$ = 4$ms$ , $T_2$ = 5$ms$ and $T_3$ = 6 $ms$ respectively. Each task has the same execution time as 2$ms$. Task settings are shown in Table \ref{tbl:servoparameter}. 

\begin{figure}
	\centering
	\resizebox{0.76\textwidth}{!}{%
		\includegraphics{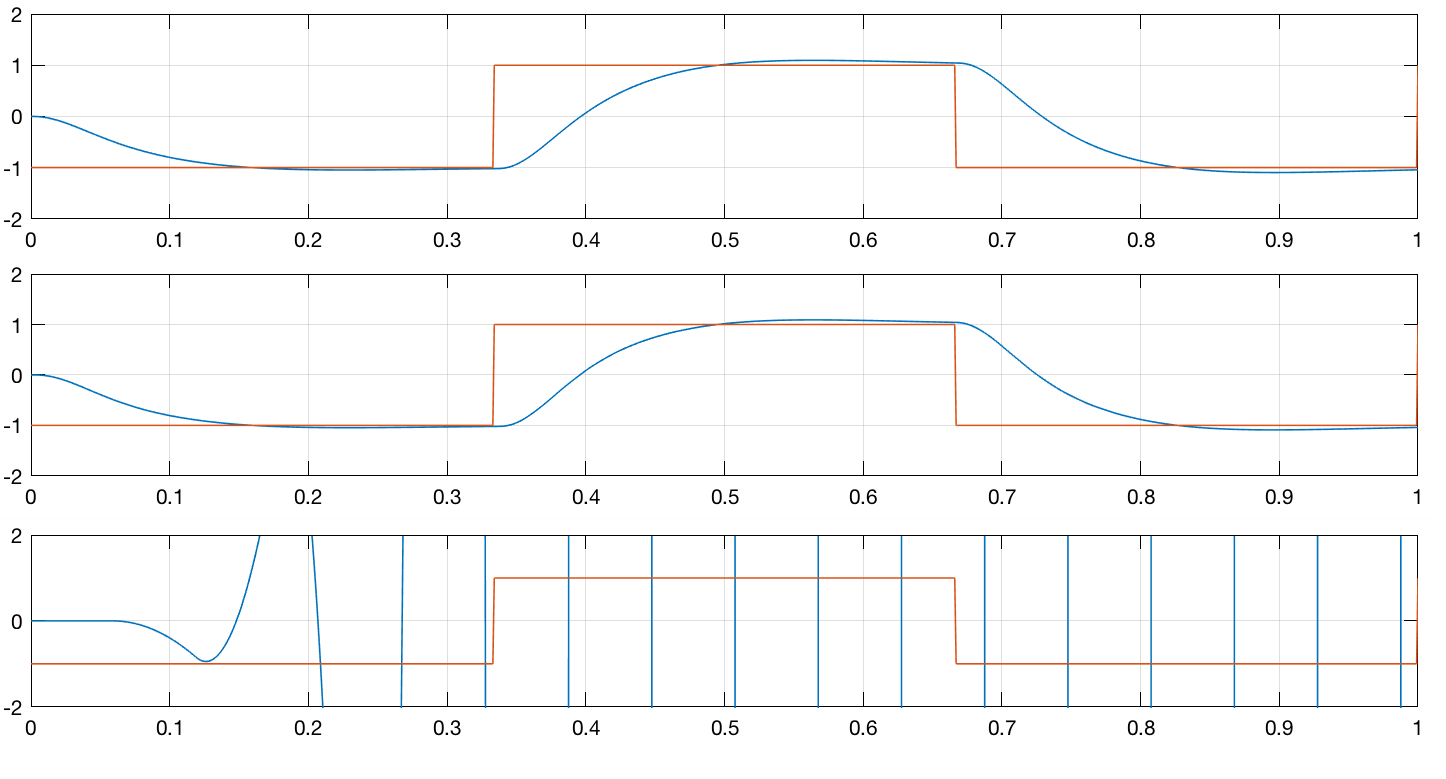} 
	}
	\caption{The output of three servos example.}
	\label{fig:MSTTTRES}
\end{figure}

\begin{figure}
	\centering
	\resizebox{0.76\textwidth}{!}{%
		\includegraphics{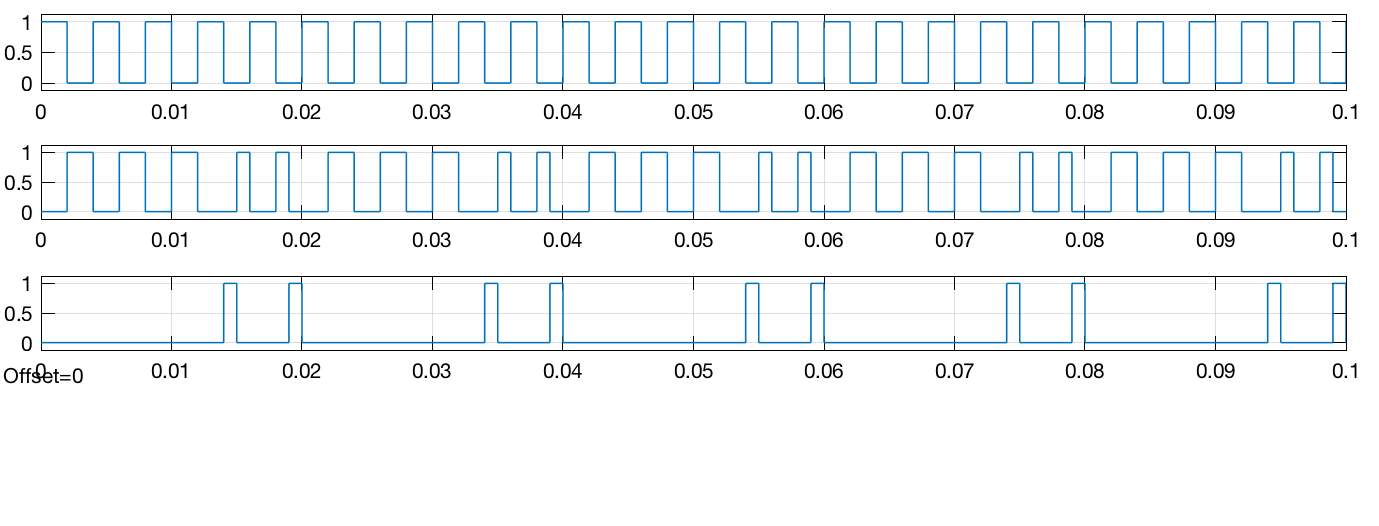} 
	}
	\caption{The task execution diagram of three servos example.}
	\label{fig:threeservoactive}
\end{figure}

Our SimSched can yield similar results to the other tools(TrueTime, TRES CPAL Controller). The output of the simulation result is shown in Figure \ref{fig:MSTTTRES}. Each output has two signals, one square wave is the reference signal for the motor, and the other one is Servos' output. The bottom output is from $T_3$, and it outputs an unstable control. Because $T_3$ has the lowest priority, it missed the deadline multiple times during the simulation. Figure \ref{fig:threeservoactive} shows the three tasks execution diagram. We can see $T_3$ misses its deadline and is preempted.
\begin{table} 
	\centering
	\caption{The simple example settings.}
	\begin{tabular}{ c c c c c}
		\hline
		Task & Period($ms$) & Execution & Priority & Runnable \\
		&                       & Time($ms$) &          &                  \\ \hline
		$T_1$   & 4 & 2 & 3 & $R_1$ \\
		$T_2$   & 5 & 2 & 2 & $R_2$ \\
		$T_2$   & 6 & 2 & 1 & $R_3$ \\
		\hline
	\end{tabular}
	\label{tbl:servoparameter}
\end{table}

\section{Limitations}
One of the shortcomings of our tool is that SimSched does not easily support single-runnable tasks. Our tool is designed to aim at AUTOSAR models and targets at the smallest executable segment runnable level. There is no runnable preemption occurrence in SimSched scheduling police due to the runnable atomic rule. Hence, if a task only contains one single runnable, SimSched cannot model task preemption at all. To cover this extreme case, we need to alter our scheduling algorithm to allow preemption among single runnable tasks. 

Since each block's execution time is an estimation, it may not accurately represent the system's real-time execution. Multiple simulations with different time parameters may need to cover the possible behaviors of the system.

\section{Future Work}
Our current model scheduler only supports periodic events. Both periodic and aperiodic tasks exist in the real-time system, and aperiodic events are necessary for automotive software. One possible area of expansion is to support aperiodic events in our model scheduler. SimSched can use an internal periodic timer to create jobs for periodic tasks. For the aperiodic tasks, we may create jobs using a separate block manually or in response to external trigger interrupts.  Further, we could add more real-time scheduling algorithms such as Earliest-deadline-first (EDF) scheduling to the model scheduler so that engineers can verify the design under different scheduling algorithms to meet diverse target platforms' requirements.

We also plan to use the modified models to identify interference between tasks. Our model scheduler currently requires execution time as a parameter to perform simulation so that we can find potential issues during the simulation phase. In the future, we perform a model scheduler simulation based on the input parameters, and we can model the execution times as variables inside the scheduler and change the value of execution until we find a potential interference.

\section{Conclusion}

SimSched can schedule Simulink models more realistically so that ML/SL simulation can reflect the real-time execution on the target platform. We implemented SimSched in an S-Function block based on the FPS algorithm written in C. It can manage the hierarchy of tasks and runnables; moreover, runnables are scheduled according to the tasks' parameters. We have demonstrated a few simple examples to show the abilities of the model scheduler. The approach discussed in this paper enables ML/SL simulation to take software execution time into account without any modification to the current models. It can fill in the gap between the semantics of model simulation and its real-time execution.

\section*{Acknowledgments}
This work is supported in part by the Natural Sciences and Engineering Research Council of Canada (NSERC).

\bibliographystyle{unsrt}  
\bibliography{bibliography}

\end{document}